\documentclass[iicol]{sn-jnl}

\usepackage{makecell}
\usepackage{array}
\usepackage{booktabs}
\usepackage{multirow}
\newcommand{\minitab}[2][l]{\begin{tabular}{#1}#2\end{tabular}}

\RequirePackage{hyperref}
\usepackage{soul}
\sethlcolor{yellow}
\usepackage{hhline}

\usepackage{siunitx}
\newcommand{\SIadj}[2]{\SI[number-unit-product={\text{-}}]{#1}{#2}}

\newcommand{\Ts}{\rule{0pt}{2.6ex}}       

\usepackage{xpatch}

\usepackage[hyperref=auto,doi=true,url=false,sorting=none,style=ext-numeric-comp,maxnames=100,backend=biber,giveninits=true,isbn=false,backref=false,date=year,articlein=false,innamebeforetitle=true,innameidem=true]{biblatex}

\renewcommand{\refname}{References}

\AtEveryBibitem{%
	\ifentrytype{book}{%
		\clearname{editor}
	    \clearfield{edition}
    	\clearfield{volume}}
	{}
	\ifentrytype{incollection}{%
		\clearfield{edition}
		\clearfield{volume}}
	{}
	\ifentrytype{inbook}{%
		\clearfield{edition}
		\clearfield{volume}}
	{}
}

\DeclareDelimFormat[bib]{editortypedelim}{\space}
\DeclareDelimFormat[bib]{innametitledelim}{\space}

\DeclareFieldFormat{editortype}{\mkbibparens{#1}}%

\DeclareFieldFormat{labelnumberwidth}{#1\adddot}

\DeclareFieldFormat[article]{pages}{#1}

\DeclareFieldFormat*{title}{#1}
\DeclareFieldFormat*[inproceedings,incollection,inbook]{booktitle}{#1}

\DeclareFieldFormat{journaltitle}{#1\isdot}

\DeclareFieldFormat[article,inbook,incollection,inproceedings,mastersthesis,thesis]{titlecase:title}{\MakeSentenceCase*{#1}}
\DeclareFieldFormat[report,techreport]{titlecase:title}{\MakeSentenceCase*{#1}}

\DeclareFieldFormat{doi}{%
	\ifhyperref{\href{https://doi.org/#1}{\nolinkurl{https://doi.org/#1}}}
	{\nolinkurl{https://doi.org/#1}}%
}

\DeclareNameAlias{default}{family-given}

\DeclareDelimFormat[bib]{nametitledelim}{\addcolon\space}

\AtEveryCitekey{\iffieldsequal{eid}{pages}{\clearfield{pages}}{}}
\AtEveryBibitem{\iffieldsequal{eid}{pages}{\clearfield{pages}}{}}

\DeclareFieldFormat[article]{number}{\mkbibparens{#1}}
\DeclareFieldFormat[article]{volume}{\mkbibbold{#1}}
\renewbibmacro*{volume+number+eid}{%
	\printfield{volume}%
	\setunit*{\addnbthinspace}%
	\printfield{number}%
	\setunit{\addcomma\space}%
	\printfield{eid}}

\renewbibmacro*{issue+date}{}
\renewbibmacro*{note+pages}{
	\printfield{note}%
	\setunit{\bibpagespunct}%
	\printfield{pages}%
	\setunit{\addspace}%
	{\printtext[parens]{%
	\usebibmacro{date}}}%
\newunit}

\renewbibmacro*{publisher+location+date}{%
	\printlist{publisher}%
	\iflistundef{location}{\setunit*{\addnbthinspace}}{\setunit*{\addcomma\space}}%
	\printlist{location}%
	\setunit*{\space}%
	{\printtext[parens]{%
	\usebibmacro{date}}}%
	\newunit}

\renewbibmacro*{institution+location+date}{%
	\printlist{institution}%
	\iflistundef{location}{\setunit*{\addnbthinspace}}{\setunit*{\addcomma\space}}%
	\printlist{location}%
	\setunit*{\space}%
	{\printtext[parens]{%
	\usebibmacro{date}}}%
	\newunit}

\xpatchbibdriver{incollection}
{\usebibmacro{chapter+pages}}
{}
{}{}

\xpatchbibdriver{inbook}
{\usebibmacro{chapter+pages}}
{}
{}{}

\renewbibmacro{in:}{%
	\ifentrytype{inproceedings}{\printunit{\newunitpunct}}{\printtext{\bibstring{in}\printunit{\intitlepunct}}}}

\newbibmacro*{incollection:maintitle+booktitle}{%
	\iftoggle{bbx:maintitleaftertitle}
	{}
	{\iffieldundef{maintitle}
		{}
		{\usebibmacro{maintitle}%
			\setunit{\maintitletitledelim}\newblock
			\iffieldundef{volume}
			{}
			{\printfield{volume}%
				\printfield{part}%
				\setunit{\voltitledelim}}}}%
	\usebibmacro{booktitle}%
	\iftoggle{bbx:maintitleaftertitle}
	{\iffieldundef{maintitle}
		{}
		{\setunit{\titlemaintitledelim}%
			\iffieldundef{volume}
			{}
			{\printfield[volumeof]{volume}%
				\printfield{part}%
				\setunit{\addspace}%
				\bibstring{ofseries}%
				\setunit{\addspace}}%
			\usebibmacro{maintitle}}}
	{}%
	\usebibmacro{chapter+pages}
	\newunit}

\xpatchbibmacro{incollection:parent}
	{\usebibmacro{maintitle+booktitle}}
	{\usebibmacro{incollection:maintitle+booktitle}}
	{}{}
	
\xpatchbibmacro{inbook:parent}
{\usebibmacro{maintitle+booktitle}}
{\usebibmacro{incollection:maintitle+booktitle}}
{}{}	
	
\newbibmacro*{inproceedings:maintitle+booktitle}{%
	\iftoggle{bbx:maintitleaftertitle}
	{}
	{\iffieldundef{maintitle}
		{}
		{\usebibmacro{maintitle}%
			\setunit{\maintitletitledelim}\newblock
			\iffieldundef{volume}
			{}
			{\printfield{volume}%
				\printfield{part}%
				\setunit{\voltitledelim}}}}%
	\usebibmacro{booktitle}%
	\iftoggle{bbx:maintitleaftertitle}
	{\iffieldundef{maintitle}
		{}
		{\setunit{\titlemaintitledelim}%
			\iffieldundef{volume}
			{}
			{\printfield[volumeof]{volume}%
				\printfield{part}%
				\setunit{\addspace}%
				\bibstring{ofseries}%
				\setunit{\addspace}}%
			\usebibmacro{maintitle}}}
	{}%
}

\newbibmacro*{inproceedings:publisher+location+note+pages+date}{%
	\printlist{publisher}%
	\iflistundef{location}{\setunit*{\addnbthinspace}}{\setunit*{\addcomma\space}}%
	\printlist{location}%
	\iffieldundef{note}
	{}
	{\setunit*{\addcomma\addspace}%
	\printfield{note}}%
	\iffieldundef{pages}
	{}
	{\setunit*{\addcomma\addspace}%
	\printfield{pages}}%
	\setunit{\addspace}%
	\printtext[parens]{%
		\usebibmacro{date}}%
	\newunit}

\renewbibmacro*{inproceedings:parent}{%
	\iftoggle{bbx:innamebeforetitle}
	{\usebibmacro{in:editor+others}%
		\setunit{\printdelim{innametitledelim}}\newblock}
	{}%
	\usebibmacro{inproceedings:maintitle+booktitle}%
	\setunit{\addcomma\addspace}
	\usebibmacro{inproceedings:publisher+location+note+pages+date}}

\xpatchbibdriver{inproceedings}
{\usebibmacro{chapter+pages}}
{}
{}{}

\jyear{2022}%

\begin{document}

\title[Explosive Launcher]{An Explosively~Driven~Launcher Capable of \SI{10}{\kilo\metre\per\second}~Projectile~Velocities}

\author*[1]{\fnm{Justin} \sur{Huneault}}\email{justin.huneault@mail.mcgill.ca}
\author*[2]{\fnm{Jason} \sur{Loiseau}}\email{jason.loiseau@rmc.ca}
\author[1]{\fnm{Myles T.} \sur{Hildebrand}}\email{myles.hildebrand@mail.mcgill.ca}
\author[1]{\fnm{Andrew J.} \sur{Higgins}}\email{andrew.higgins@mcgill.ca}

\affil*[1]{\orgdiv{Department of Mechanical Engineering}, \orgname{McGill University}, \orgaddress{\street{Macdonald Engineering Building, 817 Sherbrooke Street West}, \city{Montreal}, \postcode{H3A 0C3}, \state{QC}, \country{Canada}}}

\affil[2]{\orgdiv{Department of Chemistry and Chemical Engineering}, \orgname{Royal Military College of Canada}, \orgaddress{\street{Sawyer Building, 11 General Crerar Crescent}, \city{Kingston}, \postcode{K7K 7B4}, \state{ON}, \country{Canada}}}



\abstract{Launching large (> \SI{1}{\gram}) well-characterized projectiles to velocities beyond \SI{10}{\kilo\metre\per\second} is of interest for a number of scientific fields but is beyond the reach of current hypervelocity launcher technology. This paper reports the development of an explosively driven light-gas gun that has demonstrated the ability to launch \SIadj{8}{\milli\metre} diameter, \SI{0.36}{\gram} magnesium projectiles to \SI{10.4}{\kilo\metre\per\second}. The implosion-driven launcher (IDL) uses the linear implosion of a pressurized tube to shock-compress helium gas to a pressure of \SI{5}{\giga\pascal}, which then expands to propel a projectile to hypervelocity. The launch cycle of the IDL is explored with the use of down-bore velocimetry experiments and a quasi-one-dimensional internal ballistics solver. A detailed overview of the design of the \SIadj{8}{\milli\metre} launcher is presented, with an emphasis on the unique considerations which arise from the explosively driven propellant compression and the resulting extreme pressures and temperatures. The high average driving pressure results in a launcher that is compact, with a total length typically less than a meter. The possibility to scale the design to larger projectile sizes (\SIadj{25}{\milli\metre} diameter) is demonstrated. Finally, concepts for a modified launch cycle which may allow the IDL to reach significantly greater projectile velocities are explored conceptually and with preliminary experiments.}


\maketitle






\section{Introduction}
\label{Sec:Intro}

\begin{figure}
	\centering
	\includegraphics[width=1.0\columnwidth]{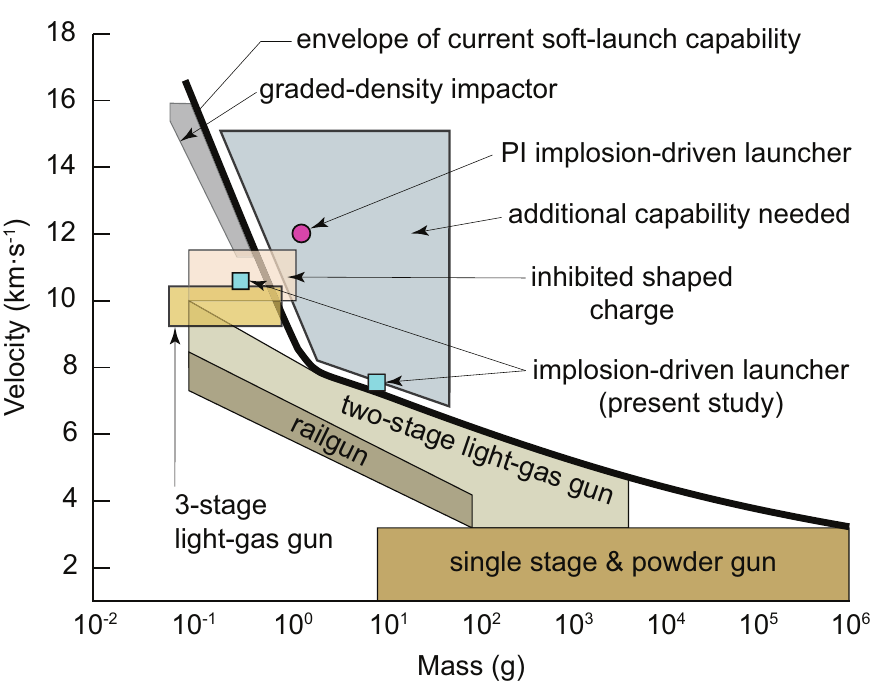}
\caption{Projectile velocities as a function of mass for widely used launcher techniques, showing the envelope of current soft launch capability. Results for the implosion-driven launcher presented in this work are included for reference. Based on \cite{NAP4765}.}
	\label{Fig:MassVelocity}
\end{figure}

The development of hypervelocity launchers based on a gasdynamic cycle have reached a velocity plateau over the entire projectile mass--velocity spectrum. The maximum projectile velocity obtainable by hypervelocity launch technologies is thus insufficient to fully meet orbital debris testing requirements (\SI{15}{\kilo\metre\per\second}) or to probe the behaviour of materials at pressures exceeding 1~TPa. The existence of the performance envelope is demonstrated by Figure~\ref{Fig:MassVelocity}, where launch velocity capability is plotted versus projectile mass for the continuum of launcher designs, ranging from conventional single-stage guns to multi-stage light-gas guns. Also included for reference are the results from the implosion-driven launcher discussed in this work. As can be seen, the ability to launch well characterized projectiles in the \SIrange{1}{10}{\gram} range to velocities above \SI{10}{\kilo\metre\per\second} lies outside the current performance envelope. This combination of projectile mass and velocity is critical to certify spacecraft shielding against orbital debris that lies below the detection threshold of ground-based radar~\cite{NAP13244}. It is also required for the study of materials at high pressure, such as planetary impact geophysics~\cite{Ahrens1987,Asimow2015,Duffy2019} and asteroid deflection~\cite{RN798}.

The two-stage light-gas gun is the workhorse of hypervelocity research for launch velocities between \SIrange{4}{10}{\kilo\metre\per\second}~\cite{Canning1970,Bogdanoff1997}. Light-gas gun velocity performance is limited by the maximum pressure that the launcher can contain and the ablation of wall material that results from the high propellant temperature and flow velocity~\cite{Bogdanoff1998}. Consequently, light-gas guns are typically operated well below their technical limits, at velocities of around \SI{7}{\kilo\metre\per\second}, in order to minimize damage to the launcher. Future optimizations promise more routine operation around \SI{10}{\kilo\metre\per\second}, but these have yet to be implemented in practice~\cite{Bogdanoff2016}. Because of the long piston travel distance and large area ratio between piston and projectile, large facilities are required to launch projectiles to hypervelocity; large-bore launchers can reach \SIrange{30}{50}{\metre} in length.

Propellant sound speed can be increased even further by adding a third stage to the launch cycle. In the three-stage light-gas gun, the high-velocity projectile from a second stage is used to shock-compress another column of light gas to greater temperature in order to accelerate a final, smaller projectile~\cite{Piekutowski2006}. In practice this configuration has exceeded \SI{10}{\kilo\metre\per\second}, although the reduced area ratio between successive stages severely limits the mass of projectile that can be launched~\cite{Friend1969}. Alternatively, a graded-density impactor can be launched from a two-stage light-gas gun onto a target flyer plate which is accelerated via the resulting quasi-isentropic impact-driven compression~\cite{Chhabildas1995}. To obtain significant velocity enhancement and prevent spalling the projectile, an impact-driven third stage is necessarily limited to launching thin flyer plates of substantially smaller mass than the original impactor~\cite{Chhabildas1995}. This technique has been used to launch a \SI{0.07}{\gram} titanium flyer plate to \SI{16}{\kilo\metre\per\second}~\cite{Chhabildas1995}.

Exceeding the current performance envelope using only gasdynamic stages is expected to cause irreversible damage to at least part of the launcher~\cite{Glenn1990,Glenn1997}. With the inevitability of launcher damage in this velocity regime, it becomes economically attractive to use a disposable launch scheme and harness the power density and driving pressures of high explosives to exceed \SI{10}{\kilo\metre\per\second}. The combination of high driving pressures and focusing effects allow explosive launching techniques to accelerate projectiles to velocities inaccessible by traditional light-gas guns. The inhibited shaped charge technique permits the launching of projectiles with a length-to-diameter ratio between 3 and 4 to \SI{11}{\kilo\metre\per\second} by interrupting the formation of a shaped charge jet with a solid die inserted at the mouth of the cone~\cite{Walker1993}. In another type of energy focusing scheme, a cascade of flyer plates uses a sequence of explosive charges backing successively thinner plates to accelerate a final flyer via an over-driven detonation~\cite{Ivanov1982}. This technique, which is limited to launching thin flyers due to spallation concerns, has been used to reach velocities beyond \SI{10}{\kilo\metre\per\second}~\cite{Ivanov1982,Batkov1997}. It is also possible to accelerate a projectile without direct contact with the explosive by using the detonation products or an intermediate gas as the driver. Examples of such an approach include the UTIAS launcher~\cite{Glass1972} and the Voitenko compressor~\cite{Voitenko1964,Voitenko1966,Voitenko1966_2,Sawle1969,Voitenko1975,Tasker2020}, where explosive detonation products or an explosively driven flyer plate are used to compress the driver gas in a hemispherical cavity. Although the hemispherical focusing used by these techniques can generate large peak gas pressures and high flow velocities, the intensity and short duration of the impulse results in an inefficient launch cycle where the projectile driving pressure quickly decays and the maximum pressure must be maintained well below theoretical limits of focusing to prevent projectile destruction. Resulting projectile velocities have been unimpressive~\cite{Glass1972}.

A more attractive option for launching well characterized projectiles with an explosive driver is to sequentially detonate a hollow cylinder of explosive. Implosion of the inner surface of the explosive produces a jet at the centerline, which in turn drives a gradual, linear cumulation of high-pressure driving gas~\cite{Merzhievskii1987}. Sequential linear implosion does not result in the same peak pressure as hemispherical or conical focusing but can produce a longer loading impulse; this is ultimately beneficial for the launching of intact projectiles to high-speed. An example of this technique is the Titov launcher, which uses a cumulative jet of detonation products or fill gas within the cylindrical hollow of a tubular charge of explosive to aerodynamically drag a projectile to hypervelocity~\cite{Titov1968,Rusakov1966,Rusakov1968,Zumennov1969,Merzhievskii1987}. Dust clouds have been accelerated to \SI{25}{\kilo\metre\per\second} and micro-projectiles have been launched to \SI{14}{\kilo\metre\per\second} with this technique, although substantial ablation of projectile mass due to hypersonic drag is unavoidable~\cite{Titov1968,Fadeenko1974,Belov2003}.

A linear implosion can also be used to compress a light driver gas, similar to the pump tube of a conventional two-stage light-gas gun. This is accomplished by using a grazing detonation in an outer cladding of high explosive to sequentially implode a thin-walled metal tube filled with helium/hydrogen. The resulting shock-driven compression can generate a propellant with a very high total enthalpy, which can be expanded to propel a projectile to hypervelocity. This type of launcher was pioneered in the 1960s by the Physics International Company (PI), who demonstrated the ability to launch \SI{2}{\gram} projectiles to velocities exceeding \SI{12}{\kilo\metre\per\second}~\cite{Crosby1967,Moore1968,Watson1970_1,Watson1970_2,Baum1973_1,Baum1973_2,Seifert1974}.

\section{Launch Cycle Overview}

\begin{figure}
	\centering
	\includegraphics[width=1.0\columnwidth]{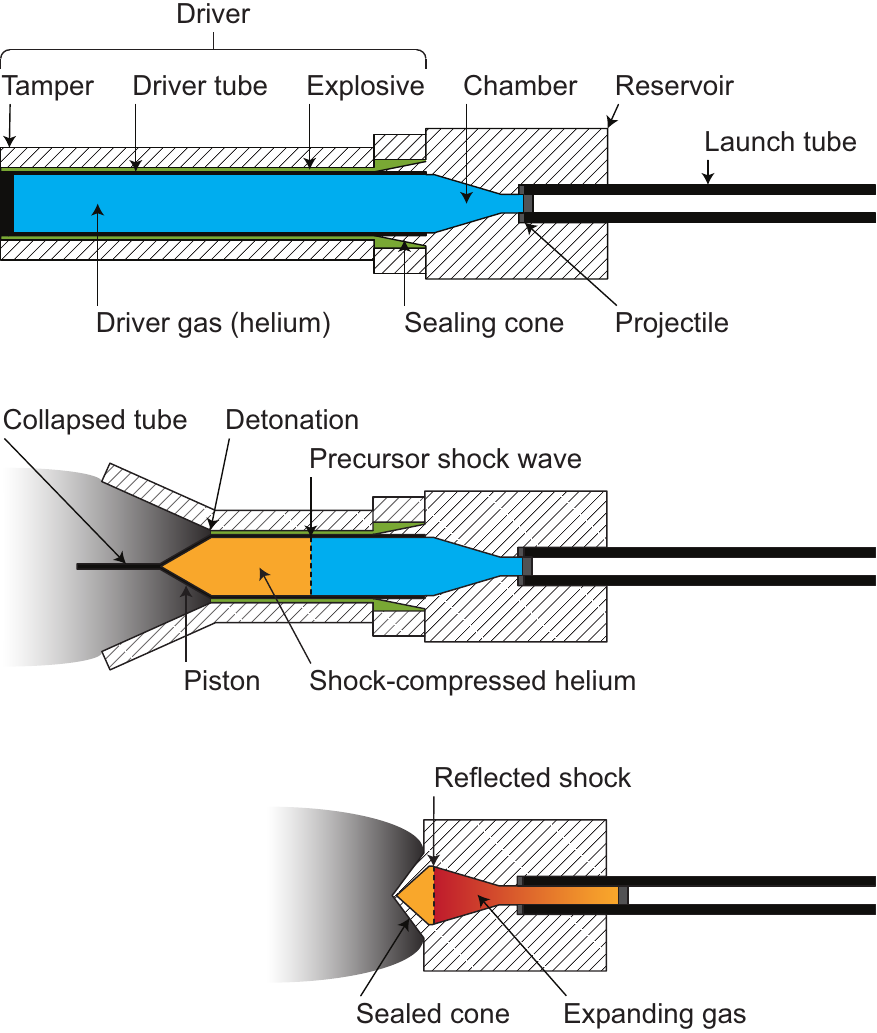}
	\caption{Overview of the launch cycle of the implosion-driven launcher.}
	\label{Fig:IGSchematic}
\end{figure}

The implosion-driven launcher (IDL), shown schematically in Figure~\ref{Fig:IGSchematic}, is essentially a light-gas gun in which the propellant is dynamically compressed by the sequential linear implosion of a pressurized tube. The driver is composed of a thin-walled steel tube pressurized with helium gas and surrounded by a thin layer of explosive, which in turn is surrounded by a thick-walled steel tamper. The explosive is initiated at the end of the tube, which causes an annular detonation wave to sweep along the tube and progressively collapse it. The imploding tube wall acts like a piston which travels at the detonation velocity of the explosive and drives a strong precursor shock wave (PSW) into the helium driver gas. As the explosively driven piston travels along the driver, an increasingly long column of shock-compressed gas is formed. The PSW travels through the chamber and reflects off the projectile, which stagnates the flow, further increasing its temperature and pressure. The reflected shock wave travels back into the shock-compressed flow, forming a reservoir of gas with a high pressure and sound speed which expands to accelerate the projectile.

The launch cycle of the IDL allows the high energy density of high explosives to be stored in a low molecular-weight gas, resulting in the launch of a projectile with much lower peak loading and to significantly greater velocities than can be reached by the direct expansion of detonation products. The linear implosion gradually focuses energy into the driver gas so that the peak projectile base pressure is less than for other explosive launcher schemes, thus allowing the launch of well characterized projectiles~\cite{Walker1993,Batkov1997,Glass1972,Sawle1969}. The explosively driven piston, which travels into the gas at the detonation velocity of the high explosive (\SIrange{5}{9}{\kilo\metre\per\second} range), causes the driver gas of the IDL to be compressed by a strong shock wave (Mach 9 for a \SI{7}{\kilo\metre\per\second} piston). As the flow is stagnated behind the projectile, the helium reaches pressures and temperatures of approximately \SI{5}{\giga\pascal} and \SI{28000}{\kelvin}. The internal energy and speed of sound obtained by strong-shock-driven compression is much greater than for the nearly isentropic compression of a typical two-stage light-gas gun pump-tube. Thus, while the IDL is conceptually similar to a gas gun, the launch cycle also shows features typical of the acceleration of a flyer plate in contact with an explosive charge because of the pressures, gas velocities, and shock waves encountered in the driver.

Since the driver is consumed by the high explosive during its operation and the helium flow properties are sufficient to cause significant plastic deformation and ablation of the chamber and launch tube, the IDL is a single-use device. However, the design of the launcher is both simple: made up primarily of stock tubing and simple machined components, and very compact: the launcher presented in this paper is less than \SI{1}{\metre} in length, resulting in modest per-shot costs. The high pressures and sound speeds that are afforded by not being limited by launcher damage and readily obtained via shock-driven compression enable the IDL to reach projectile velocities beyond the envelope of other light-gas launchers. By contrast, the use of an intermediate propellant cumulation cycle yields dramatically better piezometric efficiency~\cite{Krier_Ep,Carlucci_Ep} and propellant sound speeds compared to direct acceleration by a high explosive, which allows the IDL to actualize more of the tremendous theoretical potential of explosives.

\subsection{Explosive Driver}
\label{SubSec:ExplosiveDriver}

The explosive driver uses the sequential/linear implosion of a thin-walled steel tube to drive a normal shock wave into the helium gas. In the idealized case, the implosion pinch forms an impermeable piston that travels into the driver gas. Thus, the particle velocity behind the shock would be equal to the detonation velocity of the explosive. The precursor shock-wave (PSW) strength, described by its Mach number ($M_{\mathrm{s}}$), is then given by the following equation for a calorically perfect gas~\cite[pp. 402]{Thompson_fluids}:

\begin{equation}
\label{Equ:1}
\frac{U_{\mathrm{p}}}{c_{\mathrm{o}}}=\frac{2}{\gamma +1}\left( \frac{M_{\mathrm{s}}^{2}-1}{M_{\mathrm{s}}}  \right)
\end{equation}
\noindent 

\noindent Where $U_{\mathrm{p}}$ is piston/pinch velocity and is equal to the detonation velocity of the explosive, $c_{\mathrm{o}}$ is the initial sound speed in the gas, and $\gamma$ is the ratio of specific heats of the gas.

This equation is obtained from the normal-shock relations, which can be used to analytically determine the shock-compressed gas properties and the PSW velocity. For the optimized launcher, Primasheet explosive (63~wt\% PETN) was used, which has a detonation velocity of \SI{7}{\kilo\metre\per\second}. This results in an ideal PSW velocity of \SI{9.4}{\kilo\metre\per\second}; 34\% faster than the detonation velocity. Initial gas tube fill pressures for a typical launcher range from \SIrange{3}{7}{\mega\pascal}, resulting in post-shock pressures of \SIrange{300}{850}{\mega\pascal}. Since the PSW is continuously distancing itself from the explosive pinch, it should be possible to produce arbitrarily long columns of shocked propellant by elongating the gas tube. In practice, non-ideal effects lead to a decay in the strength of the shock wave and limit the propellant mass delivered to the chamber.

\begin{figure}
	\centering
	\includegraphics[width=1.0\columnwidth]{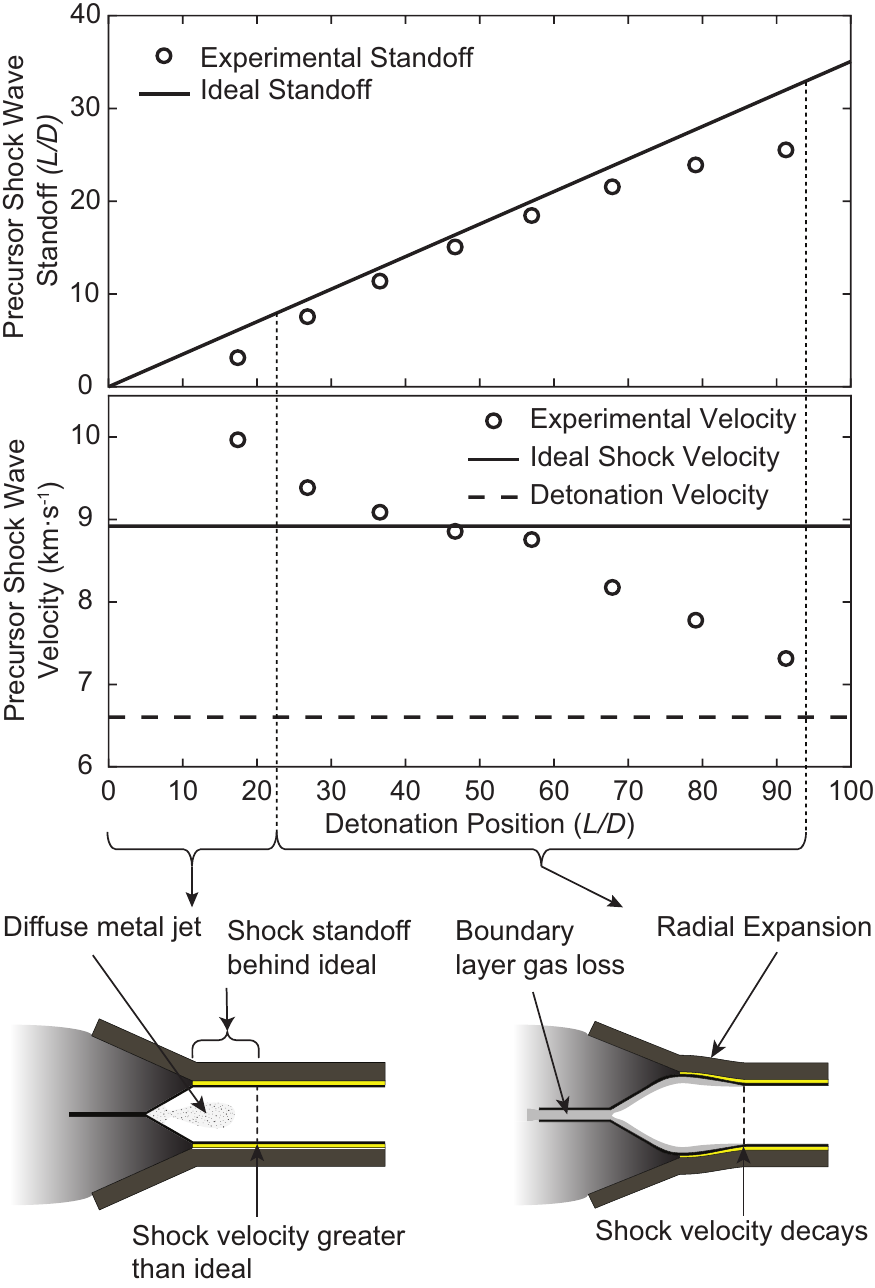}
	\caption{Overview of the non-ideal operation of the explosive driver. The precursor shock-wave standoff and velocity are plotted against the non-dimensional explosively driven piston position. The non-ideal effects responsible for observed precursor shock-wave behaviour are sketched below the plots. Results and figure adapted from Szirti et al.~\cite{Szirti2011}. The driver explosive was 90\% nitromethane sensitized with 10\% diethylenetriamine by mass; pre-compression of the liquid layer from passage of the PSW resulted in an average detonation velocity of \SI{6.6}{\kilo\metre\per\second} and an ideal PSW velocity of \SI{8.9}{\kilo\metre\per\second}. }  
	\label{Fig:NonIdealDriver}
\end{figure}

Driver non-idealities caused by the extreme gas pressures and flow conditions at the explosive pinch are depicted schematically in Figure~\ref{Fig:NonIdealDriver}. Also shown are representative graphs of the PSW standoff distance from the detonation position (i.e., the length of the compressed gas column) and PSW velocity as a function of the detonation position, which show how these parameters evolve along the length of the driver. Both the shock standoff and the detonation position are normalized by the driver-tube inner diameter. This data was obtained from experiments performed by Szirti et al.~\cite{Szirti2011} using nitromethane explosive drivers and is compared to the ideal predictions discussed above. As can be seen in top graph of Figure~\ref{Fig:NonIdealDriver}, a certain distance is required for the explosive piston to form and begin to drive a PSW ahead of the detonation, causing the standoff distance to initially lag behind ideal predictions. Approximately 10 tube diameters~\cite{Moore1968} are needed for the PSW to emerge from the collapsing tube. Beyond this start-up distance, the PSW velocity is significantly higher than ideal theory. This is the result of a cumulative jet composed of metal particulate and helium gas formed at the central implosion axis~\cite{Moore1968} and projected forward at approximately twice the detonation velocity~\cite{Walsh1953}.

Further along the driver, the PSW velocity decays from thickening of the boundary layer in the shocked driver gas and expansion of the tube walls from the intense post-shock pressure. The high-pressure gas in the boundary layer is nearly quiescent and must be driven forwards by the explosive pinch to remain in the driver. The resulting stagnation pressures are sufficiently large to prevent flow turning of the entire volume of gas trapped in the boundary layer. As a result, an increasing quantity of gas is continuously lost through the explosively driven piston, leading to a decay in the PSW velocity. Evidence for this gas loss is provided by Szirti et al., who showed that imploded pressurized tubes developed a pinhole down their lengths~\cite{Szirti2011}. As the standoff between the PSW and the detonation increases, so does the time available for the thin-walled driver tube to expand under the large pressure of the shock-compressed gas. Szirti et al. developed a computational model for the expansion of the driver tube, which showed that expansion contributes to the decay of the PSW velocity, but can be mitigated by surrounding a thin layer of explosive with a heavy steel tamper~\cite{Szirti2011}. Rapid acoustic reverberations between the tamper inner-surface and the gas tube outer-surface partially arrest gas tube expansion and prevent catastrophic bursting. The tamper has the added benefit of focusing the explosive energy inwards, thus significantly reducing the required mass of explosive to effectively form the piston. As can be seen in Figure~\ref{Fig:NonIdealDriver} from an experiment with a nitromethane driver~\cite{Szirti2011}, the combined effects of boundary layer gas loss and tube expansion lead to a significant decay in the PSW velocity, which falls well below the ideal model and eventually asymptotes to the detonation velocity, resulting in a fixed standoff distance. It should be noted that these effects become more pronounced as the initial pressure of the helium gas is increased, to the point where at sufficiently high initial pressures (typically beyond \SIrange{10}{15}{\mega\pascal}) a PSW can no longer be driven into the gas. The decay in PSW velocity, and thus shock strength, as a function of normalized driver length, leads to a significant reduction in the kinetic and internal energy of the shock-compressed gas, which introduces interesting design compromises that will be discussed in Section~\ref{SubSec:DriverDesign} of this paper.

\subsection{Launch Cycle Dynamics}
\label{SubSec:LaunchCycleDynamics}

\begin{figure}[tb]
	\centering
	\includegraphics[width=1.0\columnwidth]{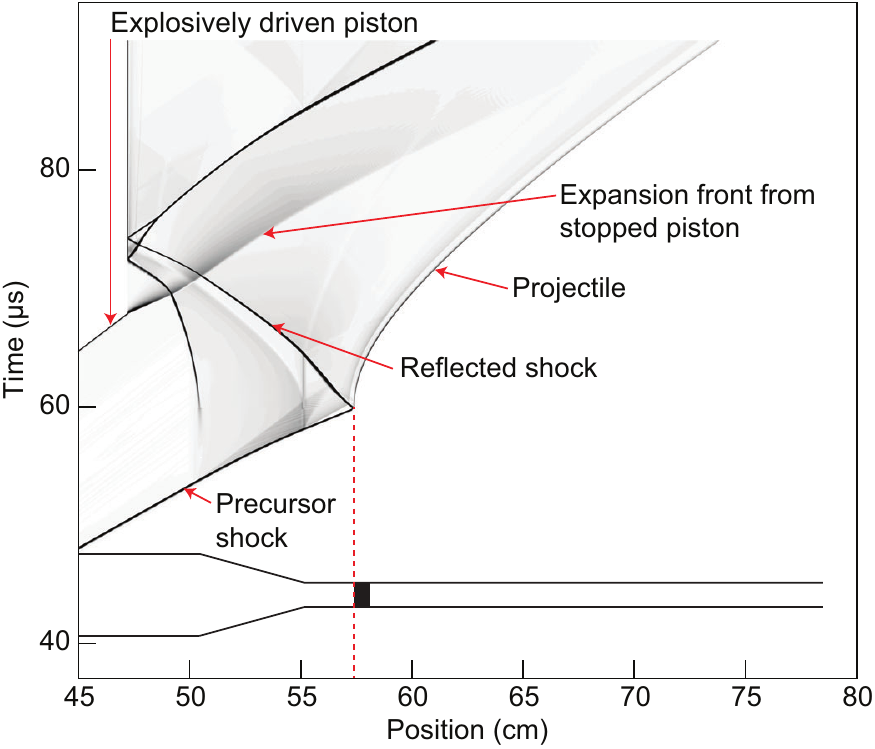}
	\caption{Labelled schlieren position--time plot showing the internal gas flow of the implosion-driven launcher. The density gradients in the flow demonstrate the main features of the launch cycle.}
	\label{Fig:Schlieren}
\end{figure}

The launch cycle of the IDL is affected by the dynamics of the shock wave as well as non-ideal effects that result from the high temperature, pressure, and flow velocity of the driver gas. The main features of the launch cycle discussed below are illustrated by the time-position schlieren plot of Figure~\ref{Fig:Schlieren}, which has been generated using Tecplot 360 to visualize an internal ballistics simulation. The schlieren image shows the density gradients in the internal gas flow, which helps identify the key wave-dynamics of the launch cycle. As the PSW reaches the end of the driver and enters the chamber, it is amplified by a gradual area reduction section that transitions from the driver diameter to the s. As a result, the peak pressure on the projectile is not only affected by the initial driver gas pressure and the strength of the PSW, but also by the ratio of the driver and launch-tube diameters, or \textit{chambrage}. The shock wave then passes through a short constant-diameter section before reaching the projectile, which allows the PSW to planarize and ensures the projectile is loaded uniformly. The peak launch cycle pressure occurs as the PSW reflects off the projectile. Although this pressure is typically much greater than the tensile strength of the projectile material, confinement from the launch tube prevents significant plastic deformation as long as the loading is uniform. Also of concern to projectile integrity is the shock wave that is transmitted into the projectile upon PSW reflection. It is well known that shock-wave interactions with free surfaces, in this case the projectile front face, can lead to tensile spall failure~\cite{Antoun2003}.

The reflection of the PSW from the projectile causes a shock wave to travel back into the chamber. This shock wave essentially stagnates the flow, leading to a significant increase in the pressure and temperature of the helium gas. The reservoir that contains the chamber and early launch tube sections is made very thick such that material strength and more importantly inertia help contain the gas pressure, which is an order of magnitude greater than the tensile strength of the wall material. The rapid expansion of the reservoir inner wall causes the ultimate projectile velocity capability of the IDL to be very sensitive to the timescale of the launch cycle: heavier projectiles, which accelerate slower, will reach significantly lower maximum velocities due to energy loss from reservoir expansion. In an attempt to minimize driver gas loss through the back of the reservoir, a tapered section is added to the end of the driver where a thick-walled “sealing cone” section is imploded. The length of the chamber section is chosen such that the reflected shock wave approaches the back of the chamber as the detonation in the explosive reaches the end of the driver and collapses the sealing cone. As the explosively driven piston effectively stops when reaching the chamber, a strong expansion front is sent into the flow. While this front causes a significant decrease in the pressure within the chamber, the launcher is designed such that the projectile has reached its maximum velocity before the expansion front reaches the projectile.

\begin{figure}
	\centering
	\includegraphics[width=1.0\columnwidth]{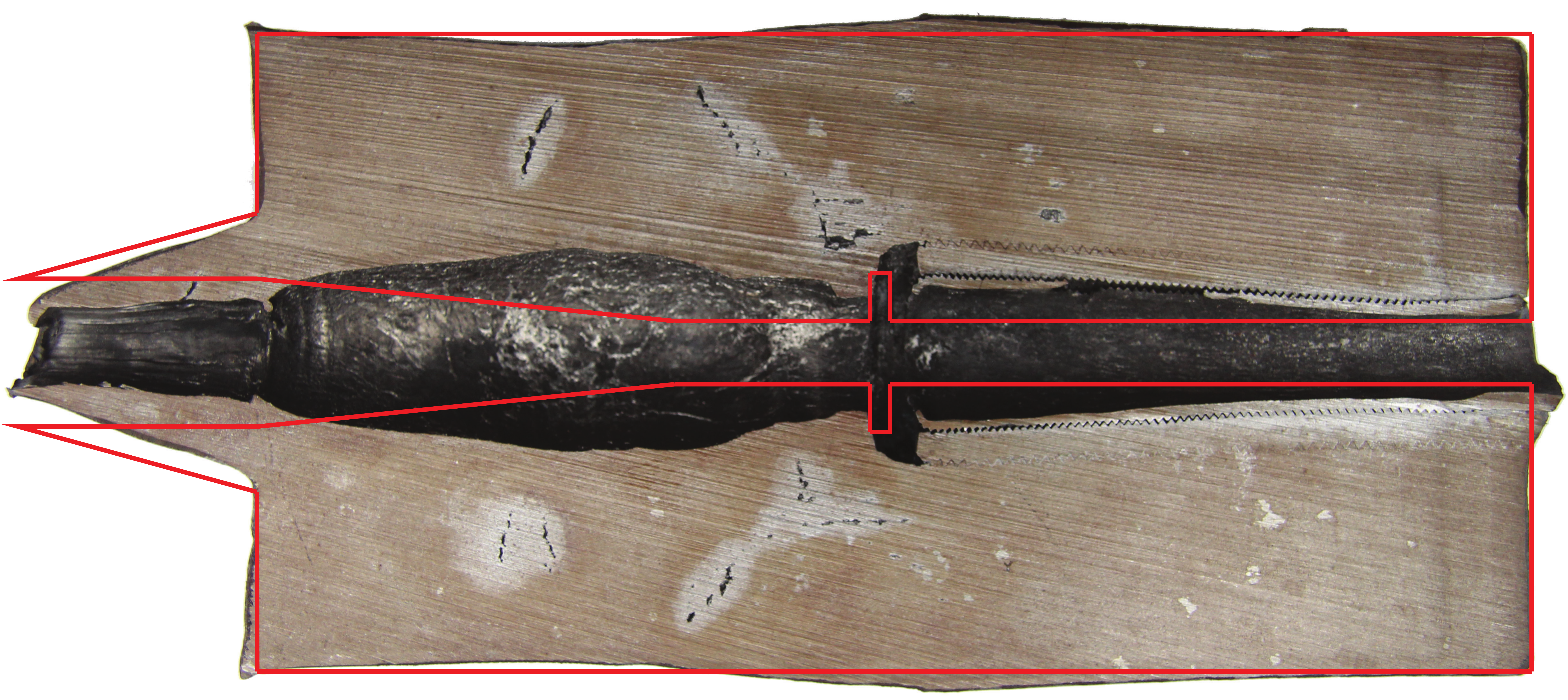}
	\caption{Photograph of a launcher reservoir that has been sectioned in half after being fired. The original profile is shown in the red outline, while the initial projectile position can be seen by the groove located approximately in the center of the reservoir. Note that this experiment included explosives on the exterior of the aft section of the reservoir.}
	\label{Fig:ReservoirSection}
\end{figure}

Throughout the launch cycle, the walls of the launcher are exposed to hot high-velocity gas flows. It has been shown that ablation can have a significant effect on the performance of two-stage light-gas guns, where the flow through the launch tube and chamber causes ablation and mixing of high-molecular-weight wall material with the light driver gas, which leads to a reduction in the maximum projectile velocity~\cite{Bogdanoff1998}. Ablation in the chamber and launch tube is an even greater concern for the IDL, where the propellant temperatures caused by the shock-driven compression are much greater than in a two-stage light-gas gun. For the IDL, ablation is also a concern in the driver, where the rapid gas flow behind the PSW can cause ablative mixing of wall material with the light gas~\cite{Hildebrand2014}, which results in the driver gas being heavily polluted before the projectile begins accelerating. Experiments performed to quantify the level of ablation by measuring the mass loss at the driver inner wall showed that the mass of ablated material is on the order of the total mass of helium gas in the driver~\cite{Hildebrand2014}. The effects of reservoir expansion and ablation will be discussed quantitatively later in the manuscript, but the importance of these effects is illustrated in Figure~\ref{Fig:ReservoirSection}, which shows a cross section of a launcher reservoir that has been cut in half after being shot. The original cross section has been traced for reference, with the initial projectile position indicated by the groove near the reservoir midpoint. The image shows severe radial expansion of the reservoir walls, especially near the original projectile position where the diameter has increased by a factor of two. It should be noted that for this test the first half of the reservoir exterior was surrounded with explosives that were detonated to mitigate reservoir expansion, which is why the final outer diameter is smaller than the original diameter in some areas. Ablative wear and deposited ablated-driver-wall material can be seen on the inner surface of the reservoir.

\section{Down-Bore Velocimetry}
\label{Sec:DownBore}

Recording the projectile acceleration history via down-bore velocimetry offers valuable insight into the launch cycle dynamics and a benchmark for internal ballistic models. The IDL is particularly well suited to performing down-bore velocimetry, as the high rate of acceleration results in relatively short projectile acceleration distances: the projectile typically reaches 80\% of maximum velocity at a travel of 10 projectile diameters. Laser interferometry techniques used in the shock physics community, such as photonic-Doppler velocimetry (PDV)~\cite{Strand2006}, have the ability to measure kilometer-per-second free-surface velocities with nanosecond temporal resolution. The insight gained from down-bore velocimetry is invaluable to the development of the IDL, since the internal ballistics of the launcher can be resolved directly in the absence of traditional diagnostic methods such as pressure transducers, whose use is precluded by the extreme launch cycle conditions.

\begin{figure*}
	\centering
	\includegraphics[width=1.0\textwidth]{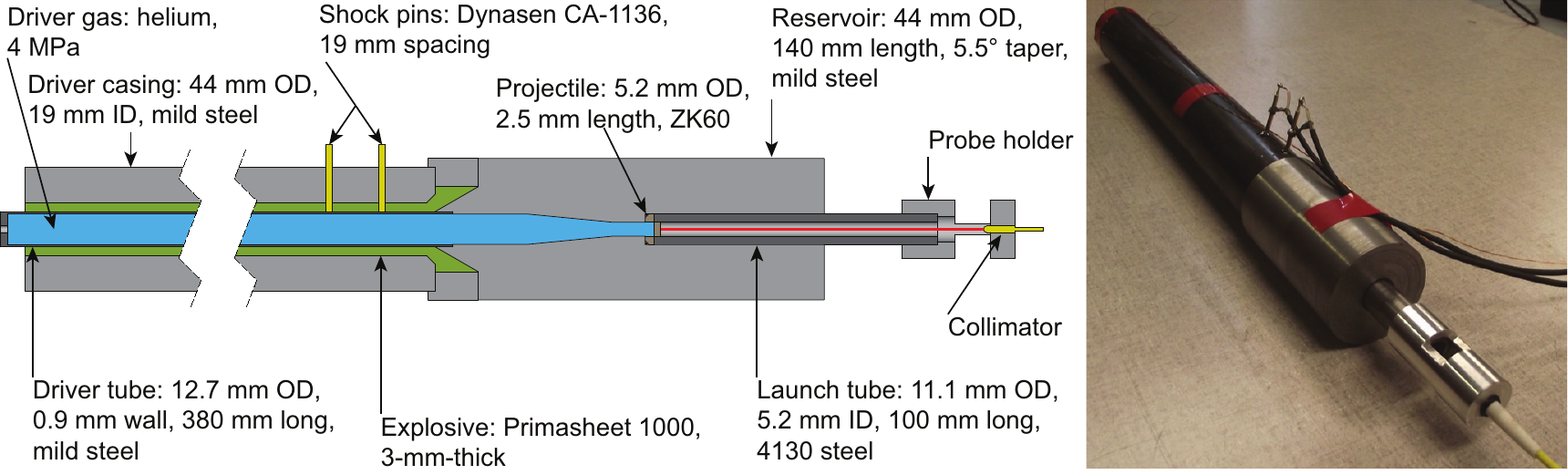}
	\caption{Schematic (left) and photograph (right) of the experimental arrangement for the down-bore velocimetry experiment.}
	\label{Fig:PDVgun}
\end{figure*}

This section will present a down-bore velocimetry experiment performed using PDV, in which the velocity history of a \SI{0.1}{\gram}, \SIadj{5}{\milli\metre}-diameter IDL projectile was tracked from 0 to \SI{7.8}{\kilo\metre\per\second}. PDV operates by blending Doppler-shifted laser light reflected from a moving surface with a source laser signal to extract a beat frequency that is proportional to the surface velocity. The PDV system uses detectors and a digitizer with bandwidths of \SI{13}{\giga\hertz}. The system was operated in a frequency-shifted mode by using a second, tunable laser as the reference~\cite{Dolan_PDV_extreme}. Offsetting the reference laser frequency relative to the source laser permits accessing higher surface velocities (in this case \textgreater{}\SI{10}{\kilo\metre\per\second}).

The experimental arrangement is shown schematically in Figure~\ref{Fig:PDVgun}. The PDV system was coupled to the launcher by threading an aluminum probe holder to the end of the launch tube, into which an optical collimator was epoxied in place. Relief ports were milled into the probe holder to allow the gas ahead of the projectile, air initially at 1~atm, to escape before reaching the probe. The launch tube length was shortened to a length of 20~projectile diameters (\SI{100}{\milli\metre}) in order to limit laser beam divergence before reaching the projectile. The launcher driver had a length of \SI{180}{\milli\metre} and a diameter of \SI{12.7}{\milli\metre}. The cylindrical projectile, which was composed of a magnesium alloy (ZK60-T5), had a diameter of \SI{5.2}{\milli\metre} and a thickness of \SI{2.5}{\milli\metre}. The operation of the driver was monitored by detonation time-of-arrival gauges at the start of the tube and by two piezoelectric shock~pins placed near the entrance to the chamber. The shock pins pass through the tamper and explosive layer to rest against the driver tube, leading to a nearly instantaneous voltage rise as the shock wave passed over their position. The PSW velocity and position can be readily determined from the pair of shock pins, while the standoff between the detonation and shock wave can be determined from the self-shorting twisted pairs that record the arrival of the detonation in the explosives and the known explosive detonation velocity. These values can be compared with ideal predictions to ensure the driver operated as expected.

\begin{figure}
	\centering
	\includegraphics[width=1.0\columnwidth]{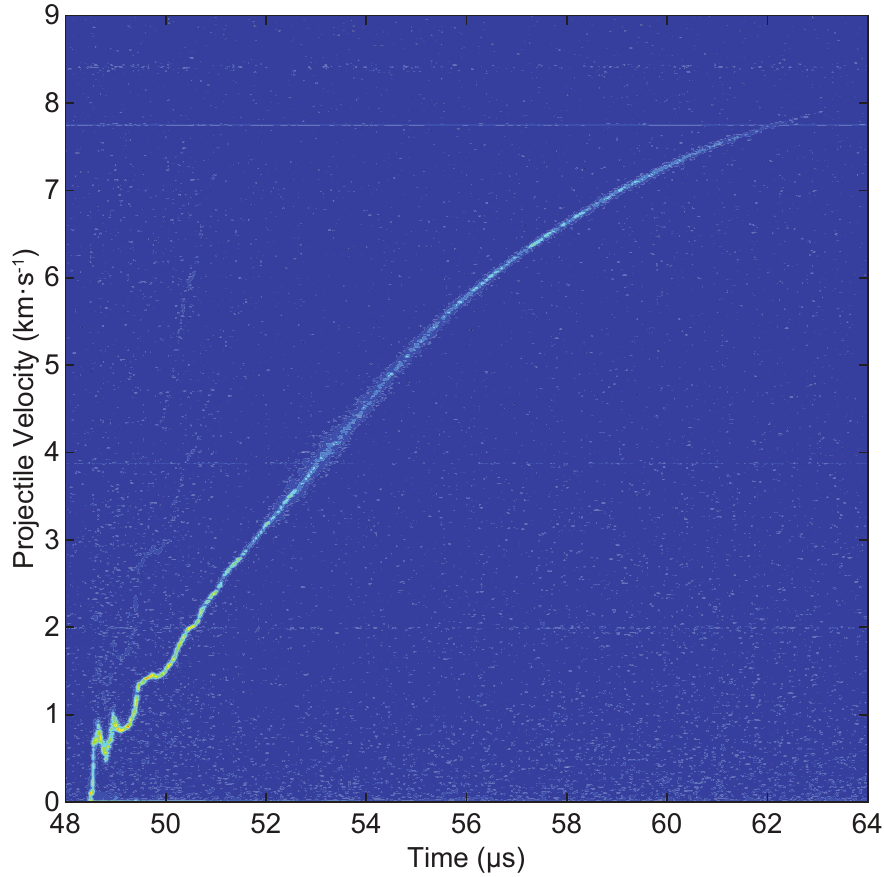}
	\caption{Photonic Doppler velocimetry spectrogram of the down-bore velocimetry experiment. The light-colored trace shows the evolution of projectile velocity as a function of time. The color contours show the power spectral density of a sliding window fast Fourier transform performed on the PDV data, where the frequency has been converted to velocity using the Doppler formula.}
	\label{Fig:DownboreSpectro}
\end{figure}

The PDV data obtained from the experiment was used to construct the spectrogram shown in Figure~\ref{Fig:DownboreSpectro}, in which the lighter portion of the contour plot indicates the projectile velocity as a function of time. It should be noted that the time axis zero for the experiment corresponds to the initiation of the detonation in the driver. The strong return of a single signal in the spectrogram is indicative of an intact projectile with smooth surface; significant ejecta or fragmentation of the projectile surface would appear as broadening of the PDV signal \cite{Georges2017}. As can be seen in Figure~\ref{Fig:DownboreSpectro}, the PDV signal was lost as the projectile reached a velocity of \SI{7.8}{\kilo\metre\per\second}, which is typical of down bore velocimetry experiments performed on IDLs, and is believed to be caused by obstruction of the laser signal by driver gas that leaks around the projectile~\cite{Hildebrand2017,HildebrandThesis}. A comparison of the shock-wave time of arrival, velocity, and standoff values at the end of the driver obtained from the shock pins with those of an ideal driver discussed in Section~\ref{SubSec:ExplosiveDriver} is shown in Table~\ref{tab:Downborespdata}. As can be seen, the experimental values were very close to ideal, as would be expected for a driver with a length-to-diameter ratio of 35.

\begin{table*}
\begin{center}
\caption{Comparison of the precursor shock-wave time of arrival, velocity, and standoff measured in the down-bore velocimetry experiment to that of an ideal driver.}
\label{tab:Downborespdata}
\tabcolsep7pt\begin{tabular}{l l l l}\toprule
Parameter 				& Measured 							& Ideal 							& \% Difference \\ \midrule
Shock arrival at pin 1 	& \SI{37.1}{\micro\second} 			& \SI{36.3}{\micro\second}         	& 2.1\% 		\\
Shock arrival at pin 2 	& \SI{39.1}{\micro\second} 			& \SI{38.3}{\micro\second}         	& 2.0\% 		\\
Shock velocity 			& \SI{9.5}{\kilo\metre\per\second} 	& \SI{9.4}{\kilo\metre\per\second} 	& 0.9\% 		\\ 
Standoff 				& 88 mm 							& 94 mm 							& 5.7\%			\\ \bottomrule
\end{tabular}
\end{center}
\end{table*}

Certain key features of the IDL launch cycle are illustrated by the down-bore velocimetry data. The shock wave transmitted to the projectile caused a nearly instantaneous jump in projectile velocity to \SI{1}{\kilo\metre\per\second}. This shock wave was followed by release waves that caused the projectile free-surface velocity to oscillate after the shock. This \textit{pullback} after the shock indicates the presence of significant tensile loading within the projectile and that spallation is an important consideration (discussed in Section~\ref{SubSec:Projectile} below). Another important feature of the launch cycle is the rapid timescale over which the projectile accelerates. The entire PDV signal lasted \SI{12}{\micro\second}, at which point the projectile had already reached 80\textendash{90\%} of its maximum velocity. It is instructive to note that arrival of the detonation at the end of the driver (\SI{54.4}{\micro\second}) corresponded to a projectile velocity of \SI{4.9}{\kilo\metre\per\second}, reinforcing the fact that the projectile launch occurs on the same timescale as the shock-wave reverberations.

\subsection{Comparison to an Internal Ballistics Model}
\label{SubSec:ModelComp}

\begin{figure}
	\centering
	\includegraphics[width=1.0\columnwidth]{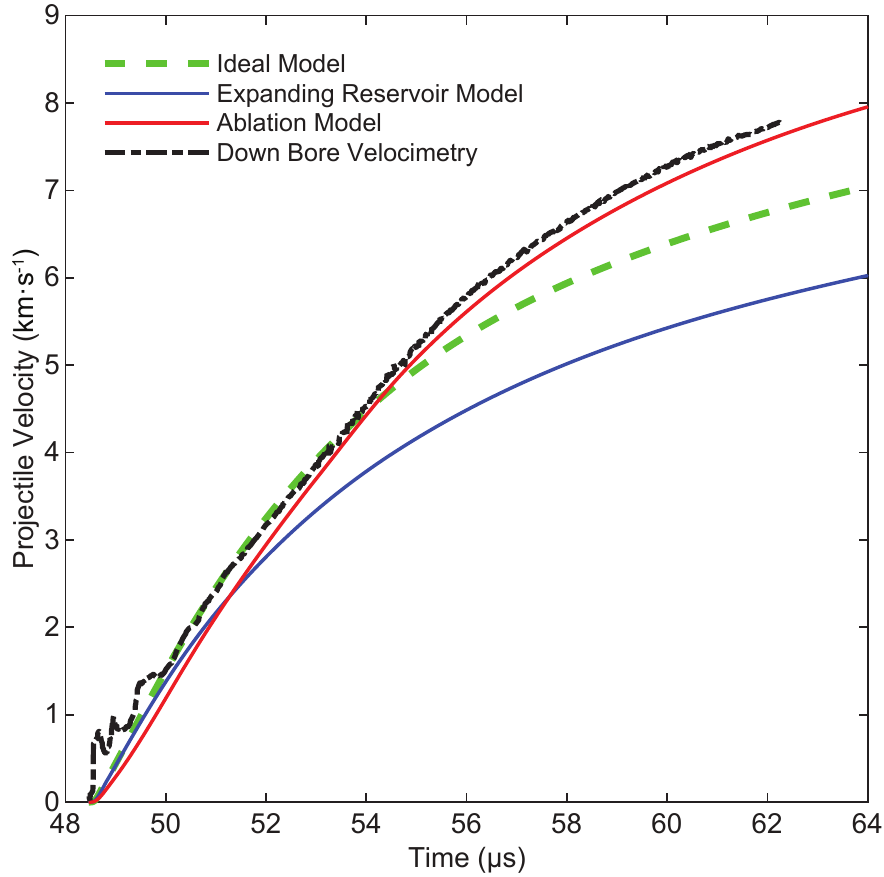}
	\caption{Projectile velocity as a function of time for the down bore velocimetry experiment. The experimental results are compared to simulations from an internal ballistics solver with different non-ideal models.}
	\label{Fig:DownboreVT}
\end{figure}

The experimental results presented above can be compared to a computational model based on a quasi-one-dimensional Internal Ballistics Solver (IBS) developed specifically to simulate the IDL~\cite{HuneaultThesis,Huneault2013}. The model is based on a second-order-accurate Lagrangian finite difference scheme with artificial viscosity developed by von Neumann and Richtmyer~\cite{VonNeumann1950}. The IBS treats the projectile as a lumped mass, meaning it cannot reproduce the initial shock in the projectile seen in the PDV data. The launcher driver is simulated by moving a planar piston into a rigid tube of helium gas at the explosive detonation velocity, which is a reasonable approach given the good agreement between the measured shock velocity and the ideal driver model shown in Table~\ref{tab:Downborespdata}. Due to the simplicity of the solver, it is possible to add different physical models for non-ideal effects present in the launch cycle~\cite{HuneaultThesis,Huneault2013}. A comparison of model results with the experimental down-bore velocimetry data can be seen in Figure~\ref{Fig:DownboreVT}. The data from the spectrogram of Figure~\ref{Fig:DownboreSpectro} has been extracted to form the “Down Bore Velocimetry” data of Figure~\ref{Fig:DownboreVT}. The “Ideal Model” curve represents a simulation where the walls of the launcher have been assumed to be rigid, which would be expected to provide an upper bound of projectile velocity. The “Expanding Reservoir Model” curve corresponds to a case where the IBS has been coupled to a radial hydrocode to simulate the radial expansion of the reservoir under the high gas pressures. As can be seen in Figure~\ref{Fig:DownboreVT}, both the ideal and expanding model significantly under-predict the projectile acceleration for most of the launch cycle, indicating that there is a non-ideal effect that is offsetting losses to the point where the driving pressure on the projectile is maintained above the ideal simulation.

Agreement between IBS simulations and the down-bore velocimetry data is improved by the addition of an ablation model in which the molecular weight of the gas is increased due to the mixing of melted wall material with the helium. The effect of ablation has been simulated using a simplified version of the model developed by Bogdanoff~\cite{Bogdanoff1998}, where the mass of melted wall material is determined by calculating the energy deposited to the launcher walls from convective heat transfer. A tuning parameter is used to correct for the fact that a portion of the ablated material will be entrained in the boundary layer and not mix with the bulk driver gas (see Section~\ref{SubSec:ExplosiveDriver}). The IBS simulation incorporating both the ablation model and the expanding reservoir model is shown in Figure~\ref{Fig:DownboreVT} as the ``Ablation Model'' curve. As can be seen, the addition of ablation to the simulation increases projectile acceleration early in the launch cycle and can be further tuned to provide good agreement with the PDV data. In typical simulations, this tuning parameter had a value of 8.4; yielding 12\% of the ablated mass mixing into the propellant through the jetting process (see Figure~\ref{Fig:NonIdealDriver}). This value is consistent with the jetted mass in acute-angle shaped charges. Despite the limitations of this relatively simple model, it provides strong evidence that ablation is the mechanism responsible for the observed projectile acceleration exceeding that of the ideal IBS simulation.

Previous ablation studies related to two-stage light-gas guns have focused on ablation in the launch tube section, where the resulting increase in the average molecular weight of the gas has a limiting effect on the projectile velocity, as the driver gas must now accelerate both the projectile and the ablated wall material~\cite{Bogdanoff1998}. The ablation present in the driver of the IDL has a significantly different effect on the launch cycle because the polluted gas has a large initial particle velocity and a non-uniform distribution along the column of compressed gas. Indeed, the increase in the molecular weight of the gas within the driver results in the explosive doing more work on the gas, which leads to an increase in the energy available to accelerate the projectile. It is also important to consider that the column of compressed gas between the explosively driven piston and the PSW is expected to contain a linearly increasing mass of ablated wall material because the gas near the back of the column has travelled a further distance along the driver tube. Thus, during the initial stages of projectile acceleration, the reflected shock wave will stagnate the increasingly heavy gas towards the back of the compressed gas column, which results in a greater-than-expected pressure behind the reflected shock wave. This increase in pressure is transmitted to the projectile by compression waves which help maintain a high driving pressure. Ablation in the driver therefore produces a mostly favorable situation in which the helium gas near the projectile, which will be accelerated to many kilometers-per-second, is initially nearly free of ablated wall material, while the gas near the back of the compressed gas column sees a significant increase in molecular weight, which increases the total energy available to accelerate the projectile. However, it should be noted that the flow of propellant in the IDL will also cause ablation in the launch tube. The additional energy needed to accelerate this entrained mass of ablated material will reduce the maximum velocity that can be achieved by the projectile and could be a limiting factor in the performance of the launcher. Further exploration of exploiting this effect by introducing a gradient of molecular weight in the gas in the driver tube---in order to achieve greater velocities---is explored in Appendix~\ref{App:Advanced}.

\section{Detailed Launcher Design}
\label{Sec:DetailedDesign}

This section will present an overview of the design of an 8-mm-bore IDL with a demonstrated capability to launch a \SI{0.36}{\gram} projectile to velocities beyond \SI{10}{\kilo\metre\per\second}. Results from two experiments performed using the \SIadj{8}{\milli\metre}-dia-bore IDL are presented in Table~\ref{tab:8mmResults}, in which the launcher was fired into an evacuated test section and observed using a Photron SA5 video camera. While the image resolution of the projectile was limited (12 pixels by 6 pixels), the projectiles appeared to be intact, with no discernible fragments. A witness plate at the end of the target chamber was also used to ensure the projectile was not fragmented. 

\begin{table}
\begin{center}
\caption{Summary of results for two launcher experiments where the velocity exceeded \SI{10}{\kilo\metre\per\second}.}
\label{tab:8mmResults}
\renewcommand{\tabcolsep}{5pt} 
\begin{tabular}{llll}\toprule
{Caliber} & {Driver Gas} & {Projectile} & {Velocity} \\ \midrule
\SI{8}{\milli\metre} & He, \SI{4.1}{\mega\pascal} & ZK60, \SI{0.36}{\gram} &  \SI{10.4}{\kilo\metre\per\second} \\ 
\SI{8}{\milli\metre} & He, \SI{4.1}{\mega\pascal} & ZK60, \SI{0.36}{\gram} &  \SI{10.2}{\kilo\metre\per\second} \\ \bottomrule
\end{tabular}
\end{center}
\end{table}

\begin{figure*}
	\centering
	\includegraphics[width=\textwidth]{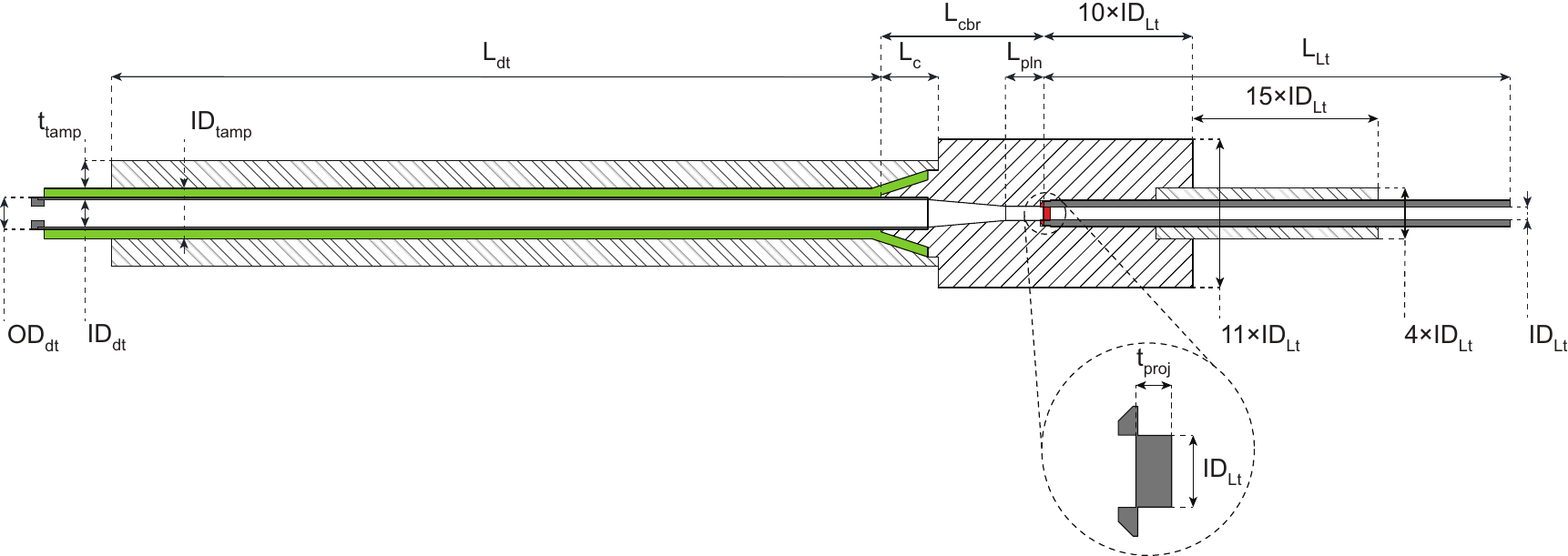}
	\caption{Overview of the design of the \SIadj{8}{\milli\metre} implosion-driven launcher. Values for the design parameters can be found in Table~\ref{tab:8mmLauncher}.}
	\label{Fig:8mmLauncherDrawing}
\end{figure*}

A schematic of the \SIadj{8}{\milli\metre} launcher is shown in Figure~\ref{Fig:8mmLauncherDrawing}, which highlights the key details and design parameters that will be discussed in the following subsections. The total length of the launcher is \SI{0.91}{\metre}. Values for the design parameters are summarized in Table~\ref{tab:8mmLauncher}. The discussion below will focus on the design considerations that result from the dynamic driver gas compression as well as the extreme propellant pressures, temperatures, and flow velocities in the IDL. The \SIadj{8}{\milli\metre} launcher design is based on using helium rather than hydrogen as the driver gas, a choice which is discussed in detail in Appendix~\ref{App:Hydrogen}. The design parameters will be discussed in their non-dimensional form in order to apply to launchers of different sizes.

\begin{table}
\begin{center}
\caption{Main design parameters of the \SIadj{8}{\milli\metre} launcher. Values are presented in non-dimensional form where possible for use in launcher scaling.}
\label{tab:8mmLauncher}
\setlength{\tabcolsep}{2pt}
\begin{tabular}{p{0.82\linewidth} p{0.18\linewidth}}\toprule
	\multicolumn{2}{ c }{\textbf{Projectile}}																												\\ \toprule
	Projectile \& Launch Tube Inner Diameter (ID\textsubscript{lt}) 							  		& \SI{8}{\milli\metre}  					\\ \cmidrule{1-2}
	Projectile Thickness to Diameter Ratio (\textit{t}\textsubscript{proj}/ID\textsubscript{lt}) 		& 0.5                   					\\ \cmidrule{1-2}
	Projectile Material 																				  		& ZK60                  					\\ \cmidrule{1-2}
	Projectile Density ($\rho_{\mathrm{proj}}$) 										 						& 1.83~g/cm\textsuperscript{3} 				\\ \cmidrule{1-2}
	Normalized Projectile Mass ($\frac{\rho_{\mathrm{proj}} t_{\mathrm{proj}}}{\mathrm{ID}_{\mathrm{lt}}}$) 	& 0.92~g/cm\textsuperscript{3}				\\ \toprule
	\multicolumn{2}{ c }{\textbf{Driver Design}}																									    	\\ \toprule
	Driver-Tube Thickness to Outer Diameter Ratio (\textit{t}\textsubscript{dt}/OD\textsubscript{dt}) 	& 0.065 			  						\\ \cmidrule{1-2}
	Explosive Thickness 									  	 	                                            & \SI{5.6}{\milli\metre} 	 				\\ 
	\cmidrule{1-2}
	Tamper Thickness to Inner Diameter Ratio (\textit{t}\textsubscript{tamp}/ID\textsubscript{tamp}) 	& 0.55 										\\ \cmidrule{1-2}
	Explosive Mass to Driver-Tube Mass Ratio (\textit{C}/\textit{M}) 											& 1.1 										\\ \toprule
	\multicolumn{2}{ c }{\textbf{Driver Geometry}}																											\\ \toprule
	Driver Length to Inner Diameter Ratio (\textit{L}\textsubscript{dt}/ID\textsubscript{dt}) 			& 28  										\\ \cmidrule{1-2}
	Chambrage Ratio (ID\textsubscript{dt}/ID\textsubscript{lt}) 								& 2.1 										\\ \cmidrule{1-2}
	Driver Fill Pressure 																						& \SI{4.1}{\mega\pascal} 					\\
	\cmidrule{1-2}
	Driver Gas Mass to Projectile Mass Ratio (\textit{G}/\textit{M}) 											& 1.9			 							\\ \toprule
	\multicolumn{2}{ c }{\textbf{Reservoir and Launch Tube}} 																								\\ \toprule
	Chamber Length to Driver Length Ratio (\textit{L}\textsubscript{cbr}/\textit{L}\textsubscript{dt}) 			& 0.22 										\\
	\cmidrule{1-2}
	Sealing Cone Length to Driver Inner Diameter Ratio (\textit{L}\textsubscript{c}/ID\textsubscript{dt}) & 2.2									\\
	\cmidrule{1-2}
	Shock Planarization Length to Projectile Diameter Ratio (\textit{L}\textsubscript{pln}/ID\textsubscript{lt}) & 2.8 							\\
	\cmidrule{1-2}
	Area Change Taper Half Angle 																				& \SI{5}{\degree} 							\\
	\cmidrule{1-2}
 	Sealing Cone Taper Half Angle 																				& \SI{18}{\degree} 							\\
	\cmidrule{1-2}
	Launch Tube Length to Inner Diameter Ratio (\textit{L}\textsubscript{lt}/ID\textsubscript{lt}) 	& 38 \\										   \bottomrule
\end{tabular}
\end{center}
\end{table}

\subsection{Projectile Considerations}
\label{SubSec:Projectile}

The projectile consists of a ZK60-T5 Magnesium alloy cylinder with a thickness (\SI{4}{\milli\metre}) equal to half its diameter (\SI{8}{\milli\metre}). The projectile thickness has been chosen to be as thin as possible in order to reduce its mass, but sufficiently thick to be stable within the bore of the launch tube. The projectile is sized such that it has a tight slip fit into the launch tube. As can be seen in Figure~\ref{Fig:8mmLauncherDrawing}, the projectile has a set of tabs held together by a thin web (\SI{0.2}{\milli\metre}). The tabs allow the projectile to seal the initial helium gas pressure with a corner o-ring, while the web is made with a sufficient thickness to hold the initial driver gas pressure. This self-sealing projectile arrangement eliminates any flow disturbances or shock interactions that would otherwise occur with a separate sealing diaphragm and is easily implemented in the IDL due to the fact that the pressure ramp up from the reflected shock wave is instantaneous, as opposed to the gradual rise in pressure encountered in two-stage light-gas guns. 

The survivability of the projectile is the limiting factor in the launch cycle of the IDL. The pressure generated by the explosive driver is not limited by the capability of the device, but rather by the fracture threshold of the projectile. Empirical evidence has shown that a reflected shock pressure of approximately \SI{5}{\giga\pascal} corresponds to the limit where the ZK60 magnesium alloy projectiles used in this study can be reliably launched intact. While the mechanism for projectile failure has not been directly observed, the two most likely failure mechanisms are uneven loading on the projectile and shock-wave interactions within the projectile and at the interface between the projectile and launch tube. While it is possible to avoid uneven loading on the projectile by ensuring concentricity of mating parts, allowing the PSW to planarize before reaching the projectile, and avoiding the use of a diaphragm before the projectile, shock interactions within the projectile cannot be entirely eliminated. The magnitude of shock-wave interactions will be proportional to the pressure behind the reflected shock wave, which explains the dependence of projectile survivability on maximum pressure. The use of a projectile material whose acoustic impedance (density multiplied by speed of sound) is closely matched to that of the steel launch tube prevents significant shock interactions at the interface between the projectile and launch tube. The projectile material should also have a sufficiently high spall strength to survive the tensile stresses caused by wave interactions. The acoustic impedance matching criteria and the need for a high spall strength precludes the use of plastics as projectile materials due to their significantly lower speed of sound and spall strength compared to lightweight metal alloys. We emphasize that the use of a self-sealing projectile (thus avoiding the use of a diaphragm that would apply uneven loading to the projectile upon rupture) is critical to operate the launcher near the limit of projectile survivability and actualize more of the theoretical potential.

The velocity potential of the IDL is particularly sensitive to the mass of the projectile due to the radial expansion of the reservoir that results from the GPa driver-gas pressures. For a fixed launcher geometry and fill pressure, the projectile acceleration is inversely proportional to its mass. Therefore, as the mass of the projectile increases, the launch cycle timescale also increases, which allows for more energy to be lost to radial expansion of the reservoir, an effect that was demonstrated with the internal ballistics model of the launcher described in Section~\ref{SubSec:ModelComp}~\cite{Loiseau2013_2}. Although the decrease in projectile acceleration can be offset by an increase in the initial pressure of the driver gas, this also results in an increase in the rate of radial expansion. In practical terms, this means that an increase in projectile mass must be accompanied by a proportionally greater increase in maximum pressure. It is for this reason that ZK60 is chosen as the projectile material over aluminum or titanium alloys which may have higher strength-to-weight ratios.

\begin{figure}
	\centering
	\includegraphics[width=1.0\columnwidth]{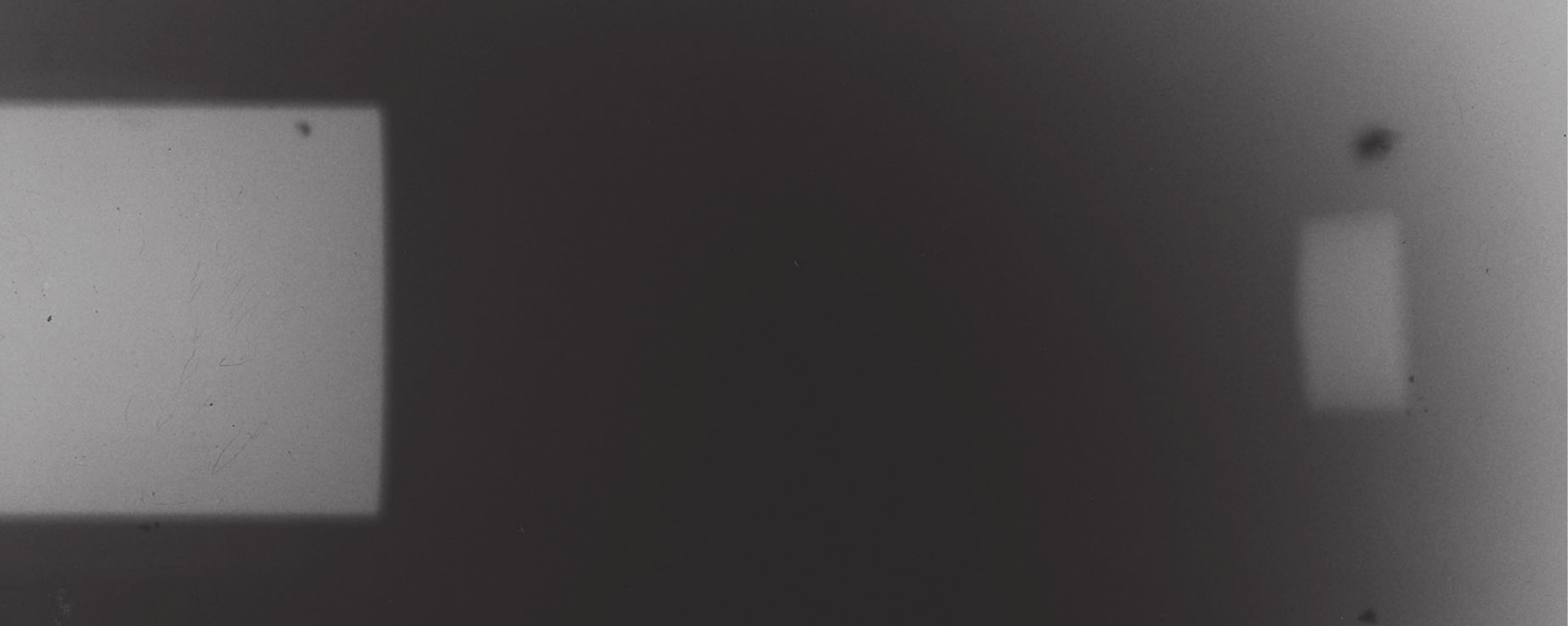}
	\caption{Flash X-ray photography of a \SI{25}{\milli\metre} projectile (right) exiting the launch tube (left) of an implosion-driven launcher.}
	\label{Fig:ProjXray}
\end{figure}

Despite the large driving pressures and shock loading that result from the launch cycle, projectiles are routinely launched intact, with no discernible sign of damage from high-speed video footage, as can be seen in the video provided in the supplemental material. Experiments have also been performed where a flash X-ray system imaged the projectile as it exited the launch tube. The use of a flash X-ray system is needed to image the projectile in the early stages of flight due to the presence of driver gas that bypasses the projectile and would obscure an optical camera. The image in Figure~\ref{Fig:ProjXray} was taken at the muzzle of a \SIadj{25}{\milli\metre}-dia-bore IDL in which the estimated reflected shock pressure was of \SI{5.3}{\giga\pascal}. The launcher, which was intended to perform projectile integrity studies, had a simplified constant area geometry, where the driver and launch tube had the same inner diameter. As a result, the peak loading is similar to that of typical IDL, but the measured projectile velocity was only \SI{5.3}{\kilo\metre\per\second}, much lower than would be achieved with a chambered launcher design. The projectile had an initial diameter (25.4~mm) equal to half the launch tube outer diameter, and a thickness (\SI{13}{\milli\metre}) of half its diameter. As can be seen, the projectile retained its original shape and was relatively planar as it exited the launch tube.

\subsection{Driver Design}
\label{SubSec:DriverDesign}
\begin{figure}
	\centering
	\includegraphics[width=1.0\columnwidth]{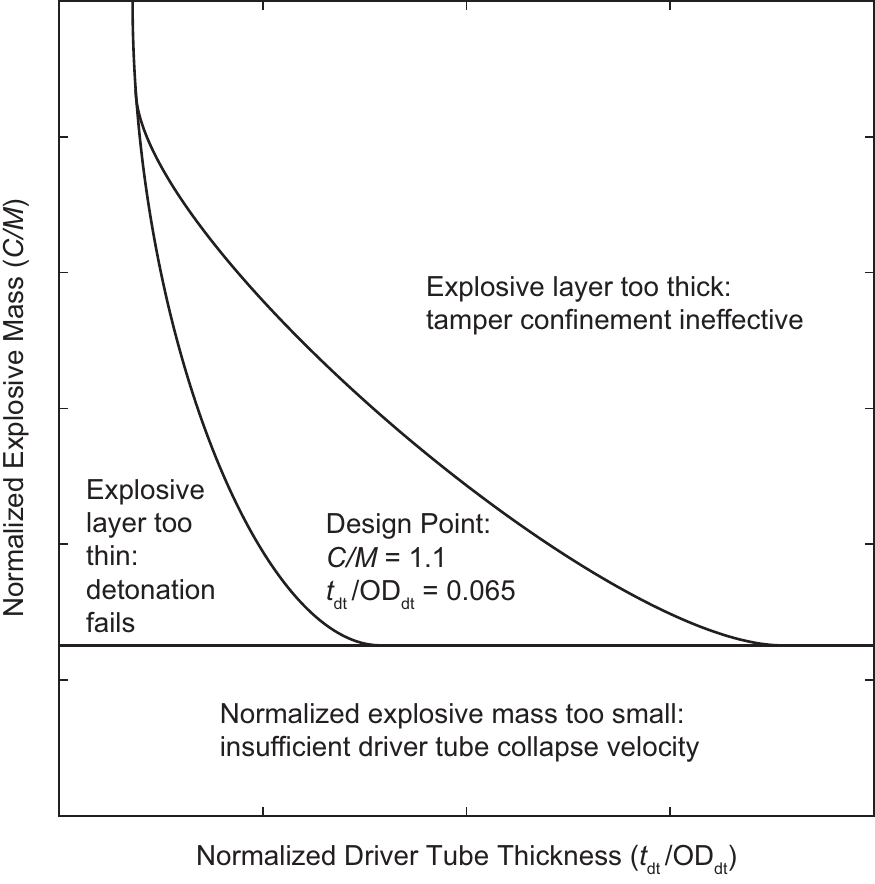}
	\caption{Parameter map of the design of the explosive driver showing the effects of varying the driver-tube thickness and the explosive mass (thickness).}
	\label{Fig:DriverDesignMap}
\end{figure}

The driver is composed of a thin-walled steel tube surrounded by a layer of explosive and a thick-walled tamper made from steel mechanical tubing. The thickness of the driver tube and explosive layer must be carefully chosen to ensure the driver operates optimally. As was discussed in Section~\ref{SubSec:ExplosiveDriver}, expansion of the driver tube can cause a significant decay in the PSW velocity, which lowers the particle velocity and enthalpy behind the shock wave and results in reduced launcher performance. The magnitude of driver-tube expansion is mainly determined by the ability of the tamper to provide confinement to the system, which requires the use of a relatively thin explosive layer~\cite{Szirti2011}. However, for the explosive driver to operate properly, the ratio of the detonation velocity of the explosive to the inward collapse velocity, essentially the aspect ratio of the implosion, must be sufficiently small in order to prevent excessive driver gas leakage through the pinch~\cite{Szirti2011}. The inward collapse velocity of the driver tube can be determined from the Gurney model, which shows that for a given explosive and tamper thickness, the wall velocity is simply a function of the ratio of the explosive mass to the driver-tube mass (\textit{C/M})~\cite{LoiseauPEP}. The implosion velocity requirement therefore results in the counter-intuitive outcome that using a \emph{thinner} driver tube, which allows for the use of a thinner explosive layer, tends to reduce radial expansion by allowing the thick-walled tamper to provide confinement to the system. The limit to how thin the driver tube and explosive layer can be made is determined by the critical thickness of the explosive, which is the limiting channel size below which the detonation wave fails to propagate~\cite{Petel2007}. If the initial explosive layer thickness is too small, expansion of the driver tube can cause the detonation to fail, which prematurely arrests the explosively driven piston and essentially terminates the launch cycle. These driver design trade-offs are illustrated qualitatively via a design parameter map  in Figure~\ref{Fig:DriverDesignMap}, where driver performance is described as a function of the driver-tube-thickness-to-outer-diameter ratio (\textit{t}\textsubscript{dt}/OD\textsubscript{dt}) and explosive-mass-to-driver-tube-mass ratio (\textit{C/M}). As can be seen, the optimal driver design uses a sufficiently thick driver tube and explosive layer to avoid detonation failure and ensure a prompt implosion. Further increasing the explosive thickness or driver-tube thickness harms the performance of the driver while simultaneously increasing the total explosive mass, which may be the limiting factor for the size of launcher that can be used at a given facility.

The driver for the \SIadj{8}{\milli\metre} launcher used an inner tube with a thickness-to-outer-diameter (\textit{t}\textsubscript{dt}/OD\textsubscript{dt}) ratio of 0.065 and an explosive thickness (\textit{t}\textsubscript{e}) corresponding to a \textit{C/M} ratio of 1.1. Designs which approximately satisfy these ratios are expected to drive a nearly ideal PSW for the driver lengths and initial helium fill pressures of interest for the IDL. The tamper is typically made of the thickest, readily available and affordable mechanical tubing, which for the \SIadj{8}{\milli\metre} launcher corresponded to a thickness-to-inner-diameter ratio (\textit{t}\textsubscript{t}/ID\textsubscript{t}) of 0.55. The explosive layer was formed by wrapping layers of Primasheet 1000, a PETN based sheet explosive whose properties are summarized in Table~\ref{tab:Explosiveproperties}. The \textit{C/M} ratio, tamper thickness, and the Gurney energy of the explosive, an empirical measurement of an explosive’s efficacy in launching flyers, determine the implosion velocity of the driver-tube wall. The \SIadj{8}{\milli\metre} launcher had a calculated implosion velocity of \SI{2.1}{\kilo\metre\per\second}, which, for a detonation velocity of \SI{7}{\kilo\metre\per\second}, gives an implosion aspect ratio of 3.3. Driver designs should seek to approximately match or exceed this value to prevent gas loss through the explosive pinch. Any explosive that can be layered, cast, poured, or pressed into a thin annulus could be used for the driver, although care must be taken to ensure gaps within the explosive assembly are minimized as this can prevent the tamper from confining the driver tube\footnote{Considerable development of the launcher reported in this study was performed with amine-sensitized liquid nitromethane as the explosive. Nitromethane was replaced with Primasheet to permit simpler evacuation of the entire chamber containing the launcher, and for launcher use on field sites that did not permit liquid explosives}. The main selection requirements for a driver explosive are that it have a high detonation velocity, density, and Gurney energy, as well as a small critical thickness. The high detonation velocity maximizes the strength of the PSW, and therefore the speed of sound of the propellant. Increasing the explosive density and Gurney energy allows for a thinner explosive layer while maintaining the desired \textit{C/M} ratio, and decreasing the critical thickness helps prevent detonation failure.

\begin{table}
\begin{center}
\caption{Properties of Primasheet 1000.}
\label{tab:Explosiveproperties}
\begin{tabular}{ l l } \toprule
Density 			&  \SI{1.44}{\gram\per\centi\meter\cubed}	\\ 
Detonation velocity &  \SI{7}{\kilo\meter\per\second}         	\\
Gurney energy 		&  \SI{2.4}{\kilo\meter\per\second} 		\\
Critical thickness 	& \textless\SI{1}{\milli\meter} 			\\ \bottomrule
\end{tabular}
\end{center}
\end{table}

\subsection{Driver Geometry}
\label{SubDriverGeo}
The diameter, length, and initial helium fill pressure of the driver determine the total amount of gas available to accelerate the projectile. nn that the ratio of driver gas mass to projectile mass (\textit{G/M}) is a critical design parameter in gasdynamic launchers~\cite{Seigel1965}. In maximizing launcher velocity potential, the goal is often to design the launcher such that it has an effectively infinite \textit{G/M}, which for the IDL means that the driver is sufficiently long such that the projectile never experiences the strong rarefaction front produced when the explosively driven piston stops. While in theory this would require an impractically long driver, reservoir expansion and ablation cause the projectile driving pressure in the IDL to decay much faster near its maximum velocity than an ideal launcher, making it possible to design the launcher such that further velocity gains from increasing driver length are negligible.

Launcher performance is also affected by the chambrage ratio, which is the ratio of the driver diameter (ID\textsubscript{dt}) to the launch-tube diameter (ID\textsubscript{lt}). For gasdynamic launchers, increasing the chambrage ratio leads to a higher average driving pressure on the projectile as the expansion front generated by the acceleration of the projectile partially reflects from the area change section as compression waves that increase the pressure acting on the projectile~\cite{Seigel1965}. For the IDL, the chambrage ratio and the geometry of the area reduction section, or chamber, that connects the driver and launch tube have an additional influence on the launch cycle because they affect the strength of the PSW and induce a relatively complex flow with significant density and velocity gradients behind the PSW~\cite{Watson1970_1}. Studies of shock-wave travel through tapered area change sections have shown that the degree of shock amplification increases with increasing area ratio and with more gradual nozzle shapes~\cite{Russell1967}; the chamber taper angle is a critical optimization parameter in light-gas guns~\cite{Bogdanoff2016}. The chambrage ratio and nozzle shape also influence the uniformity of the PSW wave when it reaches the projectile and the magnitude of density gradients in the flow behind the shock. As the PSW reflects from the projectile and travels towards the chamber, it will interact with these density gradients and cause compression waves to reflect towards the projectile, which will increase the driving pressure.

Although \textit{G/M} and chambrage considerations would typically suggest that the length (\textit{L}\textsubscript{dt}) and diameter (ID\textsubscript{dt}) of the driver should be maximized, five factors unique to the IDL must be considered in determining ideal driver geometry:

\begin{enumerate}
\item The strength of the PSW decreases with increasing normalized driver length (\textit{L}\textsubscript{dt}/ID\textsubscript{dt}) and increasing initial helium fill pressure.

\item Projectile survivability places a limit on the maximum allowable PSW strength, which therefore limits the initial driver fill pressure.

\item As the driver diameter is increased, the magnitude of PSW focusing through the chamber is increased, requiring a lower initial fill pressure to match the PSW strength at the projectile.

\item The maximum pressure of the driver gas in the chamber of a chambered launcher is not uniform due to shock focusing effects. The stagnated gas near the back of the chamber is at lower pressure than the gas near the projectile.

\item Increasing the chambrage ratio drastically increases the cost and explosive mass requirement of the IDL.
\end{enumerate}

\begin{figure}
	\centering
	\includegraphics[width=1.0\columnwidth]{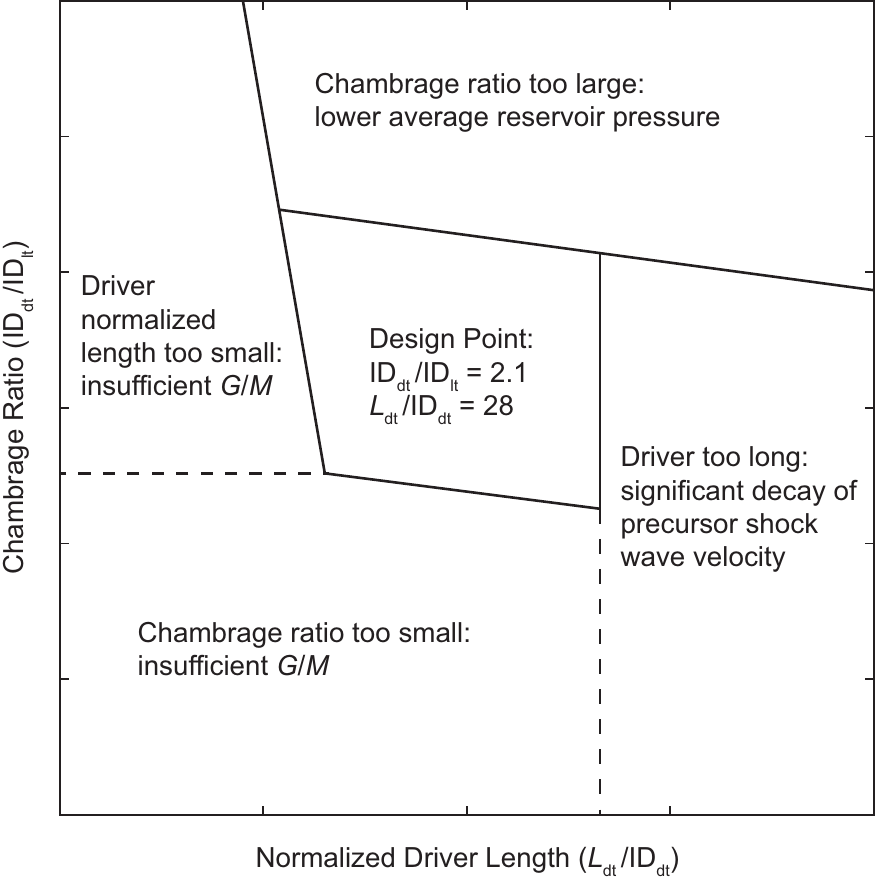}
	\caption{Parameter map showing the effect of varying the length and diameter of the explosive driver.}
	\label{Fig:DriverGeoMap}
\end{figure}

Taking these factors into consideration, the ideal driver design will have a small normalized length in order to prevent PSW decay, but a sufficiently large physical length to provide an effectively infinite \textit{G/M}. For the \SIadj{8}{\milli\metre} launcher, this is achieved by using a driver length (\textit{L}\textsubscript{dt}/ID\textsubscript{dt}) of 28 diameters, with a sufficiently large chambrage ratio (ID\textsubscript{dt}/ID\textsubscript{lt}=2.1) to reach a driver length that approaches the infinite \textit{G/M} criteria. For this geometry, an initial helium fill pressure of \SI{4}{\mega\pascal} is required to generate a reflected shock pressure of \SI{5}{\giga\pascal}, which is well within the pressure limit of the driver. A further increase in the chambrage ratio is not expected to result in a significant increase in launcher performance. This is mainly because an increase in the chambrage ratio must be accompanied by a decrease in the initial helium fill pressure of the driver to maintain a constant shock loading on the projectile. This compensatory reduction in helium fill pressure not only provides diminishing returns for increasing \textit{G/M}, but can also cause a reduction in projectile driving pressure later in the launch cycle. The maximum pressure in the IDL chamber, obtained as the flow is stagnated by the reflected shock wave, is not spatially uniform for a chambered launcher: the maximum pressure decreases with increasing diameter due to shock focusing. If the maximum pressure at the projectile is held constant for projectile integrity reasons, then the average pressure in the chamber decreases as the chambrage ratio is increased. This results in a trade-off where launchers with large chambrage ratios eventually see a faster decay in the projectile driving pressure because of the lower overall chamber pressure. The design trade-offs involved in determining the driver geometry are illustrated in a parameter map shown in Figure~\ref{Fig:DriverGeoMap}.

\begin{figure}
	\centering
	\includegraphics[width=1.0\columnwidth]{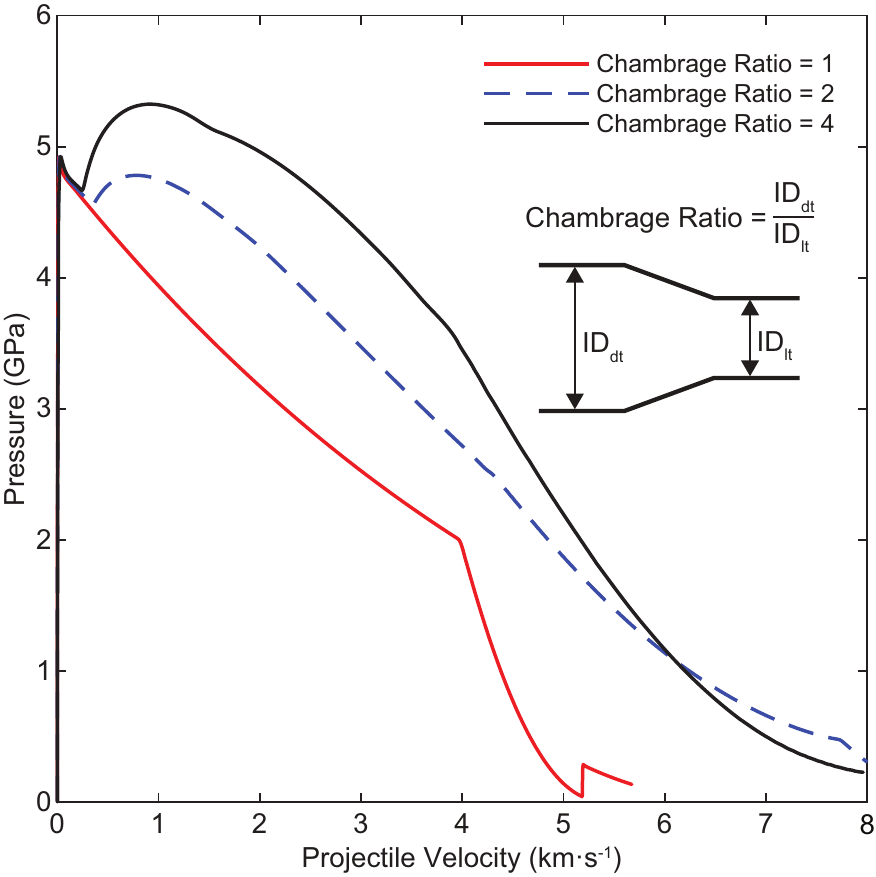}
	\caption{Projectile driving pressure as a function of velocity for internal ballistics simulations in which the chambrage ratio was varied. The length of the driver was scaled with its diameter.}
	\label{Fig:Chambrage}
\end{figure}

The effect of increasing the chambrage ratio discussed above is illustrated in Figure~\ref{Fig:Chambrage}, which shows the projectile driving pressure as a function of velocity for a series of ideal 1-D simulations, where ablation, reservoir expansion, and non-ideal driver-tube effects have been neglected. The simulations are based on the geometry of the \SIadj{8}{\milli\metre} launcher presented above, with the driver and chamber length scaled by the chambrage ratio. Launchers with higher chambrage ratios maintain a larger pressure initially, mainly due to interactions between the reflected shock wave and density gradients in the flow. The effect of not having sufficient \textit{G/M} is clearly illustrated by the constant diameter launcher (chambrage ratio=1), where a steep decrease in the projectile driving pressure is seen beyond \SI{4}{\kilo\metre\per\second} due to the arrival of the expansion front originating from the stopped driver piston. The effect of reduced chamber pressures at high values of chambrage is slightly more subtle, but is illustrated by the fact that in the range of \SIrange{4}{7}{\kilo\metre\per\second}, the projectile driving pressure decays much faster at the highest value of chambrage than in less chambered launchers. It should be noted that the optimum driver normalized length and chambrage chosen for the \SIadj{8}{\milli\metre} launcher are specific to the normalized mass ($\frac{\rho_{\mathrm{proj}} t_{\mathrm{proj}}}{ID_{\mathrm{lt}}}$) of the projectile used in this study. It should be expected that an increase in normalized mass, caused either by higher projectile material density ($\rho_{\mathrm{proj}}$) or thickness ($t_{\mathrm{proj}}$), would be accompanied by an increase in optimum driver size.

The area reduction section between the driver and launch tube is made gradual (\SI{5}{\degree} straight taper) to minimize flow disturbances behind the PSW. As was discussed in Section~\ref{SubSec:Projectile}, a length of 2.8 launch-tube diameters was given between the end of the area change section and the back of the projectile to allow the PSW to planarize before reaching the projectile. Choosing this length was a compromise between ensuring uniform loading on the projectile and maximizing projectile velocity by ensuring a short communication time between the projectile and the area change section. The effect of the internal chamber geometry has not been studied computationally, which would require a two-dimensional solver, or experimentally, and it may be possible that it could be optimized to maintain a satisfactorily uniform loading on the projectile while increasing the driving pressure through the interaction of the reflected shock wave with density gradients in the flow. The authors of the present study largely adopted the convergence angles and planarization lengths established by PI~\cite{Watson1970_1}.

\subsection{Reservoir and Launch Tube}
\label{SubSecRes}
The reservoir is a thick-walled steel vessel which transitions between the driver and the launch tube and is designed to contain the high-pressure driver gas during projectile acceleration. The driver tube is welded to the reservoir, while the launch tube is threaded in to allow the projectile to be inserted and the sealing o-ring to be compressed. The length of the chamber ($L_{\mathrm{cbr}}$), which corresponds to the distance between the end of the driver and the initial position of the projectile, is chosen such that the explosively driven piston and the reflected shock wave reach the start of the chamber, or equivalently the end of the driver, at approximately the same time. This timing criteria, which is illustrated schematically on the third image of Figure~\ref{Fig:IGSchematic}, is dictated by the ratio of the chamber length to the driver length (\textit{L}\textsubscript{cbr}/\textit{L}\textsubscript{dt}), which was set to 0.22 for the \SIadj{8}{\milli\metre} launcher. If the chamber is made longer, the driver compression will end prematurely, forming an expansion front that will decrease the chamber pressure. Conversely, if the chamber is made shorter, the reflected shock will enter the driver tube, which can disrupt the explosively driven piston and again cause a strong expansion front to originate from the back of the chamber due to the loss of driver gas. Although it may seem possible to improve launcher performance by reducing the chamber length in an attempt to further compress the driver gas, similar to increasing the compression ratio in a two-stage light-gas gun, the pressures required to drive the explosively driven piston into the stagnated flow behind the reflected shock wave are much greater than explosive detonation pressures, which makes such an approach seemingly non-viable. The back of the reservoir contains a tapered conical section with an increased explosive mass thickness, which implodes to form a plug that minimizes the loss of driver gas from the back of the reservoir as the explosively driven piston reaches the end of the driver. This “sealing cone” section (see Figure~\ref{Fig:IGSchematic}), which has a length of 2.2 internal driver diameters and a taper angle of \SI{18}{\degree}, also provides a gradual transition from the very thin driver tube to the thick-walled reservoir, which reduces the likelihood of tube fracture and gas loss at the interface between the driver tube and the reservoir as the tube expands due to the high-pressure gas behind the PSW.
 
The driver gas pressures are sufficient to yield the reservoir, which must be made sufficiently thick such that its inertia provides confinement and prevents fracture within the timescale of projectile launch. A ratio of reservoir outer diameter to projectile diameter of 11 has been used for the \SIadj{8}{\milli\metre} launcher and is a reasonable design guideline. The reservoir must also extend sufficiently far ahead of the projectile to confine the launch tube until the driving pressure has decayed. The reservoir of the \SIadj{8}{\milli\metre} launcher extends 10 launch-tube inner diameters past the initial projectile position. A launch tube sleeve, which slides onto the launch tube, is added to provide additional confinement while minimizing the cost (material and machining time) of the reservoir. The sleeve has a diameter of four~projectile~diameters and extends 25~projectile~diameters ahead of the initial projectile position.

Attempts have been made to mitigate reservoir expansion and even re-compress the driver gas by surrounding the reservoir by a thick layer of explosives. To meaningfully influence projectile maximum velocity, these explosives must be detonated early in the launch cycle, typically even before the projectile begins to accelerate. In doing so, a shock wave is driven into the steel reservoir that can fracture the projectile. Although modest velocity gains have been demonstrated using reservoir explosives~\cite{Loiseau2013_2}, the additional risk to projectile integrity and the significant increase in explosive mass requirements (typically many times the explosive mass required for the driver) has precluded the use of reservoir explosives with the \SIadj{8}{\milli\metre} launcher. Nevertheless, PI demonstrated an increase of \SI{2}{\kilo\meter\per\second} (from \SI{8}{\kilo\meter\per\second} to \SI{10.2}{\kilo\meter\per\second}) using reservoir explosives~\cite{Watson1970_1}. More recently Wang et al. have demonstrated a \SI{1}{\kilo\meter\per\second} increase (from \SI{10.4}{\kilo\meter\per\second} to \SI{11.4}{\kilo\meter\per\second}) using the launcher described in the present study as a baseline~\cite{Wang2019,Wang2020,Wang2021}.

The launch tube is made from 4130 mechanical tubing having a length of 38 projectile diameters that has been honed smooth. The launch cycle of the IDL is sufficiently dynamic that further increases in launch tube length do not improve launcher performance and may actually decrease the exit velocity if the launch tube has not been evacuated~\cite{Huneault2015}. While evacuating the launch tube is common practice for light-gas gun facilities, the use of explosives in the IDL launch cycle can make it challenging to evacuate the launch tube and target chamber, as most explosive testing facilities are either outdoors, or in large indoor blast chambers that are difficult to evacuate. Nonetheless, it is highly recommended that the launch tube and target chamber be evacuated, either by evacuating the entire blast chamber or fitting the launch tube to a one-time-use sealed target assembly which can be evacuated. A compromise solution adopted through much of the development program in this paper used a helium flush at 1~atm to remove residual air from the launch tube and target chamber.

\section{Discussion}

The above sections presented a detailed overview of the launch cycle and design considerations of an explosively driven light-gas launcher capable of reaching velocities currently inaccessible to other gasdynamic launchers. Of particular interest is the fact that the compact \SIadj{8}{\milli\metre} launcher presented in this study could readily be scaled-up to launch more massive projectiles. It is well known that gasdynamic launchers should be expected to reach nearly the same projectile velocity regardless of size, as long as all proportions of the launcher are uniformly scaled~\cite[p.~376]{Cooper1996}. Using either simulations or geometric scaling, it is readily demonstrated that the ideal launch cycle of the IDL scales perfectly, as long as the normalized size of the driver, reservoir, launch tube, and projectile are held constant~\cite{Loiseau2013_1,HuneaultThesis}. The effect of launcher size on non-ideal effects must also be considered. Simple analytical treatments and simulations have been used to show that the effect of radial expansion of the reservoir on the launch cycle is independent of scale~\cite{HuneaultThesis}. While the timescale of projectile acceleration increases proportionally to the diameter of the projectile, allowing more time for radial expansion to occur, the relative rate of change in volume from radial expansion is inversely proportional to launcher size. Similarly, driver non-ideal effects (PSW formation distance, driver-tube expansion, and boundary layer growth, see Section~\ref{SubSec:ExplosiveDriver}) are also expected to be unaffected by driver size, a fact that was experimentally verified during an extensive experimental campaign undertaken by the Physics International Company which compared the PSW velocity and standoff of a \SIadj{3.5}{\centi\metre}-diameter driver with that of a \SIadj{200}{\centi\metre}-driver~\cite{Watson1970_1,Watson1970_2}. Losses that depend on surface area, such as gas friction, heat transfer, and ablation become less significant as the launcher size is increased due to the decrease in the surface-area-to-volume ratio. Of particular interest is ablation which plays a significant role in the launch cycle of the IDL, as was shown in Section~\ref{Sec:DownBore}. While ablation in the driver leads to an increase in projectile driving pressure early in the launch cycle, ablation in the launch tube increases the molecular weight of the driver gas and is likely an important factor in limiting the velocity potential of the IDL.
 
The scaling considerations discussed above indicate that the projectile velocity capability of the IDL is generally independent of launcher size, and that larger launchers may be expected to outperform the \SIadj{8}{\milli\metre} results presented above due to a reduction in surface area losses. Tests with older generation designs of the IDL have been performed on launchers ranging from \SI{5}{\milli\metre} (\SI{0.1}{\gram} projectile mass) to \SI{25}{\milli\metre} (\SI{15}{\gram} projectile mass), demonstrating velocities of \SI{9.5}{\kilo\metre\per\second} and \SI{7.6}{\kilo\metre\per\second} respectively~\cite{Loiseau2013_2,Huneault2015}. The variation in projectile velocity between these launchers and the \SIadj{8}{\milli\metre} launcher presented in this work can be attributed to evolutions in the design rather than effects of scale. The authors are confident that the \SIadj{8}{\milli\metre} design, capable of achieving \SI{10.4}{\kilo\metre\per\second} projectile velocity as presented in Section~\ref{Sec:DetailedDesign}, could be successfully scaled to launch projectiles of \SI{25}{\milli\metre} or even larger to the same velocity. Unfortunately, lack of access to outdoor ranges necessitated by the mass of explosive required when scaling-up has precluded our testing of large-caliber launchers since 2011 (when the \SI{7.6}{\kilo\metre\per\second} was demonstrated with an earlier, non-optimized design). The \SIadj{8}{\milli\metre} IDL design can be used as a baseline to estimate the requirements of large-scale launchers. Table~\ref {tab:Scaling} presents an overview of expected explosive mass and launcher size for different projectile sizes, as well as an indication of the facility required to operate the launcher.

\begin{table*}
\begin{center}
\caption{Predicted launcher size and explosive mass requirements for larger scale implosion-driven launchers obtained by scaling the 8-mm design.}
\label{tab:Scaling}
\tabcolsep7pt\begin{tabular}{cccccc}
\hline
Projectile & Projectile & Explosive & Launcher & Launcher & \multirow{3}{*}{\minitab[c]{Facility \\ Requirements}} \Ts\\
Diameter & Mass & Mass & Mass & Length & \\
(mm) & (g) & (kg) & (kg) & (m) & \\
\hline
8  & 0.36 & 0.24  & 18   & 0.84 & small indoor facility   \Ts\\
12 & 1.2  & 0.82  & 62   & 1.3  & large indoor facility   \\
16 & 2.9  & 1.90  & 150   & 1.7  & small outdoor test site \\
25 & 11.0 & 7.40  & 560  & 2.6  & large outdoor range     \\
50 & 90.0 & 59.00 & 4500 & 5.3  & large outdoor range     \\
\hline
\end{tabular}
\end{center}
\end{table*}  

The numerous design parameters and the significant cost and preparation time in developing new launchers has made it difficult to conduct parametric studies of the IDL to experimentally determine the effect of design changes on the projectile exit velocity. Additionally, conclusions drawn from parametric studies performed in one area of the parameter space do not necessarily apply when changes are made to key parameters such as the normalized projectile mass. Although the development of the IDL may appear to lend itself to hydrocode simulation, numerous challenges exist that complicate the modelling of the launch cycle. Some examples include: the converging geometry of the explosively driven piston leads to singularities in two-dimensional axisymmetric simulations, modelling ablation and the resulting two-phase flow, modelling of boundary layer growth and gas loss in the driver, modelling of projectile fracture, etc. For these reasons, the development approach taken in this work has been to focus on launcher experiments guided by a simple one-dimensional internal ballistics model to which physical models for non-ideal effects were added. The design presented in Section~\ref{Sec:DetailedDesign} is the result of numerous design iterations, and while the entire design parameter space has not been explored experimentally, it seems unlikely that significant velocity increments could be made by varying the design of the driver or the launcher geometry.

Significant improvements in the projectile velocity of the IDL may be possible by changing the way in which the explosive driver operates. In the \SIadj{8}{\milli\metre} launcher design presented above, the driver is operated at a helium fill pressure that is approximately three times lower than the experimentally determined limit due to projectile integrity concerns. This indicates that there is unused potential in the driver that may be harnessed if the launch cycle can be designed to provide additional compression to the propellant without increasing the loading on the projectile.

The launch cycle presented in this work provides no additional compression to the driver gas once the projectile has been set in motion, which results in a rapid monotonic pressure decay and a relatively inefficient launch cycle. However, it is possible to maintain a high driving pressure on the projectile in a gasdynamic launcher by continuously increasing propellant compression after the projectile has begun to accelerate, as is accomplished with the \textit{accelerated reservoir} technique of conventional two-stage light-gas guns~\cite{Charters1987}. Simulations have been used to demonstrate that when the explosively driven piston is accelerated near the end of the driver of the IDL, compression waves travel into the chamber and increase the pressure such that a nearly constant driving pressure can be maintained on the projectile~\cite{Huneault2014}. Accelerating the explosively driven piston, which can be accomplished using phase velocity methods such as a wave shaper or an explosive lens~\cite{LoiseauThesis, Loiseau2012_SW,Loiseau2014,Huneault2014}, poses a significant challenge as any loss of driver gas will offset the benefit of the increased piston velocity. Another possibility is to design the driver such that there is an initial density discontinuity in gas, either by varying the pressure or molecular weight of the propellant~\cite{HildebrandThesis}. The gas near the projectile is initially held at a lower density, while the gas at the back of the driver is at a higher density that will result in the driver operating close to its pressure limit. The density discontinuity forms a high impedance contact surface that travels into the chamber. When reflected shock waves originating from the area change section and the projectile reach the contact surface, they are partially reflected as shock waves that travel towards the projectile, further compressing the gas. With this approach, the driver can be designed such that the initial projectile shock-loading is the same or even lower than with the standard launcher, but ramps up due to the aforementioned shock interactions. The velocity benefits that result from the higher density gas at the back of the driver are based on the same effect that was seen in Section~\ref{SubSec:ModelComp}, where ablation in the driver leads to higher than anticipated projectile acceleration. Simulations indicate that successful implementation of the density discontinuity launch cycle may allow for a significant (many km/s) increase in projectile velocity. This technique is discussed in more detail in Appendix~\ref{App:Advanced}.

\section{Conclusion}

This work has demonstrated that the implosion-driven launcher (IDL) has the potential to address the need for hypervelocity testing beyond \SI{10}{\kilo\metre\per\second}. The extreme projectile loading and strong sensitivity to projectile mass mean that the IDL will likely never be a viable option for launching the types of elaborate aeroballistic packages that can be used with two-stage light-gas guns. However, tests requiring simpler projectiles, such as impact testing (orbital debris, geophysics, asteroid deflection) or flyer plate-based equation of state testing are well suited to the IDL. The projectile launched from the IDL could also be combined with the density-gradient impactor technique~\cite{Chhabildas1995} to in-turn drive thinner projectiles to even greater velocities, or be used to effect quasi-isentropic compression to previously inaccessible densities and pressures. Moreover, the use of precision-timed explosives (low jitter) and the disposable nature of the launcher offer interesting opportunities that would otherwise require massive capital investments, such as using two counter-firing launchers to double the impact velocity, or performing one-off tests of larger launchers. While the material cost of the IDL is quite low, the cost of the explosive facility and technical expertise needed to perform the testing is substantial. However, where these facilities and expertise exist, the IDL offers a very accessible avenue to performing hypervelocity impact testing. Given the velocity capability demonstrated by the IDL over a short and limited development period, as well as the apparent inevitability that any launcher routinely reaching velocities well above \SI{10}{\kilo\metre\per\second} will be, at least in part, non-reusable, further development of the IDL may offer the most promising avenue to reaching velocities well above \SI{10}{\kilo\metre\per\second} with large (\SIrange{1}{10}{\gram}), well characterized projectiles.

\backmatter

\bmhead{Acknowledgments}	
The authors would like to thank Vincent Tanguay, Daniel Szirti, Matthew Serge, and Patrick Bachelor for their help in conducting experiments and useful input during the early development of the launcher. This work was supported by the Canadian Space Agency (CSA) [Contracts No. 9F028-064201/A and 64/7012003]; Defense Research and Development Canada (DRDC) [Contract No. W7701-82047]; the Fonds de Recherche du Quebec Nature et Technologies (FRQNT) [Grant 2012-PR-148696]; and the Natural Sciences and Engineering Research Council of Canada (NSERC) [Grant 2014-06258].

\bmhead{Data Availability}
Data included in this paper is available upon reasonable request.

\printbibliography

\newcounter{docfigcount}
\setcounter{docfigcount}{\value{figure}}

\newcounter{doctblcount}
\setcounter{doctblcount}{\value{table}}

\begin{appendices}
	
\renewcommand{\thesection}{\arabic{section}}
\renewcommand{\thefigure}{\arabic{figure}}
\renewcommand{\thetable}{\arabic{table}}

\setcounter{figure}{\value{docfigcount}}
\setcounter{table}{\value{doctblcount}}        

\section{Hydrogen as a Driver Gas}
\label{App:Hydrogen}
The choice of propellant for the IDL is somewhat more nuanced than for two-stage light-gas guns, where using hydrogen instead of helium provides a significant advantage due to reduced launcher wear from ablation~\cite{Charters1987}. The lower ablation rates that result from using hydrogen allow the launcher to be operated at greater compression ratios, resulting in higher launch cycle pressures and propellant sound speed. Although launcher wear is not a concern for the IDL, the effect of driver gas choice on ablation rates must still be considered due to the fact that mixing of the heavy ablated wall material with the light gas can lead to significant reductions in the speed of sound of the propellant and ultimately limit the velocity potential of the launcher.

The rate of ablation in a hypervelocity launcher is determined by the rate of convective heat transfer to the launcher walls from the hot driver gas flow. Using Reynold’s analogy~\cite{Bogdanoff1995}, it is possible to relate the rate of convective heat transfer ($\dot{Q}$) to the frictional fluid shear force ($F_\mathrm{r}$) at the launcher walls, the flow velocity ($u$) and the difference between the total flow enthalpy ($h+\frac{1}{2}u^2$) and the static enthalpy at the wall ($h_\mathrm{w}$):

\begin{equation}
\label{Equ:2}
\dot{Q}=\frac{F_\mathrm{r}}{u} \left( h+\frac{1}{2}u^2-h_\mathrm{w} \right)
\end{equation}

\noindent The fluid shear force can be expressed in terms of the wall surface area ($A_\mathrm{w}$), the density ($\rho$), the flow velocity ($u$), and the skin friction coefficient ($C_f$):

\begin{equation}
\label{Equ:3}
F_\mathrm{r}=\frac{1}{2}A_\mathrm{w}C_f\rho u^2
\end{equation}

\noindent The rate of heat transfer per unit mass ($\dot{q}$) of gas can then be expressed as follows, where $D$ is the diameter of the channel:

\begin{equation}
\label{Equ:4}
\dot{q}=\frac{2uC_f}{D}\left(h+\frac{1}{2}u^2-h_\mathrm{w}\right)
\end{equation}

\noindent As can be seen from Equation~\ref{Equ:4}, for similar flow velocities (i.e., comparable projectile velocities) the key fluid parameter that determines the rate of heat transfer is the difference between the static flow enthalpy ($h$) and the wall enthalpy ($h_\mathrm{w}$). In a two-stage light-gas gun, the lower specific-heat ratio of hydrogen ($\gamma_{\mathrm{H_2}}=1.4$, $\gamma_{\mathrm{He}}=1.667$) as well as dissociation of hydrogen molecules make it possible to maintain maximum propellant temperatures below or near the melting point of steel ($\approx$~\SI{1720}{\kelvin}), whereas comparable helium launch cycles would generate propellant temperatures beyond \SI{6000}{\kelvin}~\cite{Glenn1997}. The advantage of hydrogen as a two-stage light-gas gun propellant can mainly be attributed to the fact that the enthalpy of hydrogen at \SI{1720}{\kelvin} is much greater than that of helium at the same temperature.

For the IDL, maximum launch cycle temperatures, estimated from ideal launcher simulations with the SESAME tabulated equation of state~\cite{SESAME_1992}, in both hydrogen (\SI{6500}{\kelvin}) and helium (\SI{27 800}{\kelvin}) are significantly above the melting point of steel~\cite{HuneaultThesis}. As a result, the wall enthalpy, which plays an important role in determining the rate of heat transfer in the two-stage light-gas gun has little effect at IDL propellant conditions because it is negligible compared to the total enthalpy of the flow for both helium and hydrogen. Table~\ref{tab:EnthalpyComp} gives a comparison of the difference between the wall enthalpy and the enthalpy behind the reflected shock wave in a helium and hydrogen IDL using the SESAME equation of state to capture real gas effects~\cite{SESAME_1992}. Despite having a much lower reflected shock temperature, the static enthalpy difference for hydrogen is only 24\% smaller than for helium. The considerable contribution of flow velocity to the total flow enthalpy will further reduce the proportional difference between ablation rates in helium and hydrogen launchers. This simple analysis indicates that while ablation rates are predicted to be larger with a helium propellant, the difference may not be as significant as might be expected and might be sufficiently small such that it is not the determining factor in the choice of propellant.

\begin{table*}
\begin{center}
	\caption{Comparison of the difference between the reflected shock enthalpy and the wall enthalpy for helium and hydrogen launchers with matched reflected shock pressures.}
	\label{tab:EnthalpyComp}
	\setlength{\tabcolsep}{5pt}
	\begin{tabular}{ l c c c c}\toprule
		\multirow{2}{*}{Driver Gas} & \makecell{Reflected Shock \\Temperature} & \makecell{Reflected Shock \\ Enthalpy} & Wall Enthalpy	& Enthalpy Difference \\
				& (K)	& (\SI{}{\mega\joule\per\kilo\gram})	& (\SI{}{\mega\joule\per\kilo\gram})	& (\SI{}{\mega\joule\per\kilo\gram}) \\ \midrule
He, \SI{4.0}{\mega\pascal}                 & 27800 & 179 & 9.3  & 170\\
H\textsubscript{2}, \SI{5.7}{\mega\pascal} & 6500  & 158 & 28.6 & 129\\ \bottomrule
\end{tabular}
\end{center}
\end{table*}  

\begin{table*}
	\begin{center}
		\caption{Summary of experiments performed to determine the effect of the driver gas on the performance of the implosion-driven launcher.}
		\label{tab:HydrogenExp}
		\setlength{\tabcolsep}{5pt}
		\begin{tabular}{lcccccc}\toprule
			Driver Gas & \makecell{Driver \\ Length \\ (\textit{L}\textsubscript{dt}/ID\textsubscript{dt})} & \makecell{Chambrage \\ Ratio \\ (ID\textsubscript{dt}/ID\textsubscript{lt})} & \makecell{Area Change\\ Taper} & \makecell{Launch Tube \\ Inner Diameter} & Projectile & Velocity \\ \midrule
			He, \SI{3.5}{\mega\pascal} & 28 & 2.3 & \SI{4.5}{\degree} & \SI{8.5}{\milli\metre} & Al7075, \SI{0.68}{\gram}  & \SI{6.8}{\kilo\metre\per\second} \\
			H\textsubscript{2}, \SI{5.4}{\mega\pascal} & 28 & 2.3 & \SI{6.5}{\degree} & \SI{8.5}{\milli\metre} & Al7075, \SI{0.68}{\gram}  & \SI{5.7}{\kilo\metre\per\second}\\
			\bottomrule
		\end{tabular}
	\end{center}
\end{table*} 

\begin{figure}
	\centering
	\includegraphics[width=1.0\columnwidth]{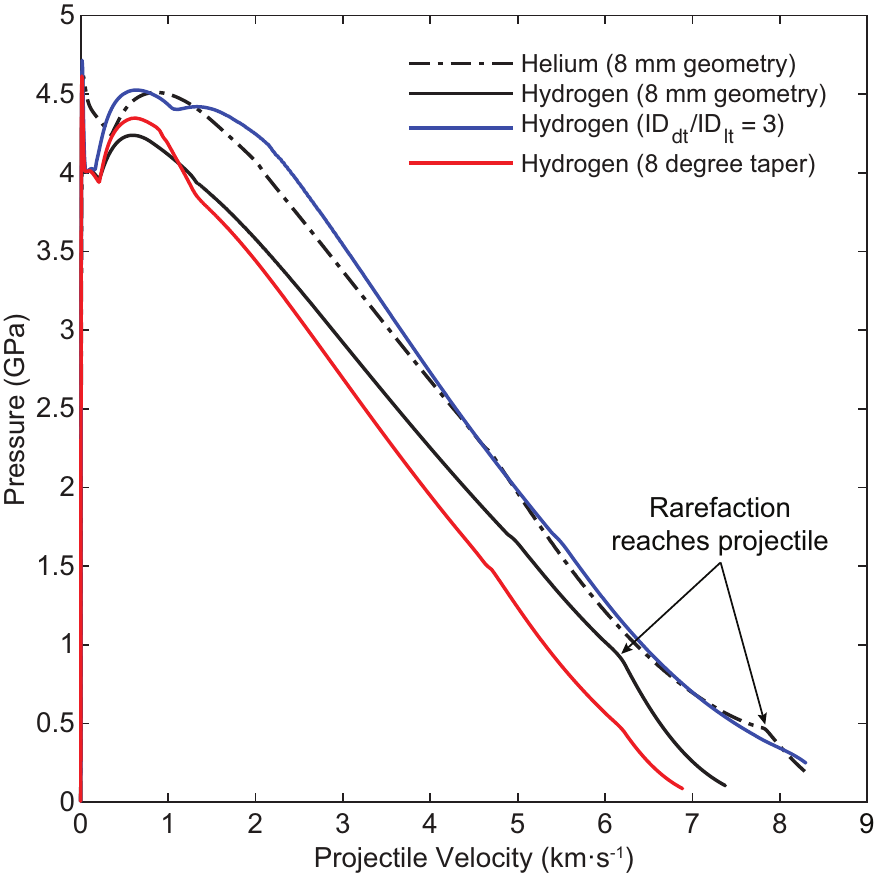}
	\caption{Projectile driving pressure as a function of velocity for internal ballistics simulations in which the \SIadj{8}{\milli\metre} helium-driven launcher is compared to hydrogen-driven launchers with different driver and chamber geometries.}
	\label{Fig:HydrogenPU}
\end{figure}

Consideration must be given to the effect of hydrogen on the launch cycle of the IDL. Ideal IDL simulations show that for a properly sized driver and matching reflected shock pressures, helium and hydrogen launchers should be expected to reach similar velocities. This is shown in Figure~\ref{Fig:HydrogenPU}, where the projectile driving pressure as a function of velocity is plotted for identical helium and hydrogen launchers. The \SIadj{8}{\milli\metre} launcher design was used for the simulations, while the driver fill pressure was adjusted to obtain matching reflected shock pressures. The ideal simulations (impermeable piston, rigid launcher walls, no ablation) used the SESAME equation of state~\cite{SESAME_1992} for both helium and hydrogen.

The most significant difference between hydrogen and helium in the IDL launch cycle is the effect of the specific-heat ratio on the shock-wave dynamics (see Equation~\ref{Equ:1}). The use of hydrogen, which has a lower specific-heat ratio and lower molecular weight, results in a reduced precursor shock-wave velocity (PSW) and pressure increase across the shock. As a result, a hydrogen launcher requires a higher initial driver fill pressure to reach the same reflected shock pressure as a helium launcher. Ideal simulations also indicate that in order to meet the effectively infinite \textit{G/M} criteria discussed in Section~\ref{SubDriverGeo}, a hydrogen launcher requires a larger driver. This can be seen in Figure~\ref{Fig:HydrogenPU}, where the inflection point from the stopped piston arrives significantly earlier for the hydrogen launcher (\SI{6.2}{\kilo\metre\per\second}) than with helium (\SI{7.8}{\kilo\metre\per\second}). Also shown in Figure~\ref{Fig:HydrogenPU} is the driving pressure as a function of velocity for a hydrogen launcher with a larger driver: chambrage ratio (ID\textsubscript{dt}/ID\textsubscript{lt}) of 3, normalized driver length (\textit{L}\textsubscript{dt}/ID\textsubscript{dt}) of 28. The significant improvement in projectile driving pressure as a function of velocity that results from the increase in driver size indicates that the hydrogen IDL requires a larger driver than a comparable helium launcher in order to maximize projectile velocity.

The lower PSW velocity in the hydrogen propellant also affects the choice of chamber length, because the shock timing criteria discussed in Section~\ref{SubSecRes} (the chamber length is chosen such that the explosively driven piston reaches the transition section between the driver and the chamber at the same time as the reflected shock wave) would require a chamber length that is less than half the length of a comparable helium chamber. However, the increase in the area change angle that would result from shortening the chamber may have a significant effect on the PSW as it travels through the area change section, influencing density gradients in the flow and the uniformity of projectile loading. The effect of shortening the chamber is shown in Figure~\ref{Fig:HydrogenPU}, where the “\SI{8}{\degree} taper” curve corresponds to a simulation where the taper angle was increased from \SI{5}{\degree} to \SI{8}{\degree} to meet the shock timing criteria. As can be seen, shortening the chamber does not appear to be beneficial, and projectile driving pressure is improved by keeping a longer area change section which does not meet the shock timing criteria, but maintains a very gradual taper. This effect would need to be further studied either experimentally or with two-dimensional simulations in order to properly capture the shock interactions in the area change section.

The results of two early launcher experiments which were performed to compare the performance of helium and hydrogen in the IDL are presented in Table~\ref{tab:HydrogenExp}. The two launchers had the same driver geometry, as well as an identical projectile and launch tube. The chamber of the hydrogen launcher was made shorter to meet the shock timing criteria discussed above. As a result, the area change taper angle was of \SI{4.5}{\degree} for the helium launcher and \SI{6.5}{\degree} for the hydrogen launcher. The driver fill pressure was chosen such that both launchers would have a similar reflected shock pressure. As can be seen, the projectile velocity from the hydrogen launcher was \SI{1.2}{\kilo\metre\per\second} (17\%) slower than for the helium launcher. The projectile velocities are much lower than those of current launcher designs, mainly from changes in the projectile material and design, as well as reducing the launch tube length and evacuating the launch tube and projectile flight path. Although it is certainly possible that the hydrogen launcher’s velocity deficit may be overcome by changes in design, notably by increasing the chambrage and the chamber taper angle, it seems unlikely that a hydrogen launcher would significantly surpass velocities obtained with helium propellant. The authors acknowledge that it may be of interest to re-visit the comparison given the recent improvements in launcher design and understanding of the launch cycle.

\section{Advanced Launch Cycle Concepts}
\label{App:Advanced}
A significant increase in the maximum projectile velocity of the IDL may be possible with modifications to the launch cycle such that additional compression is provided to the driver gas during projectile acceleration. This was recognized in the initial development of the IDL by Physics International, where a technique was proposed in which a second explosively driven piston would collapse the launch tube behind the projectile~\cite{Moore1968,Watson1970_1}. The explosively driven pinch would be accelerated using explosive lensing techniques such that it kept a constant standoff with the projectile, thus preventing the driver gas from expanding. Correctly setting the timing and rate of the launch tube implosion as to obtain a meaningful velocity increase without damaging the projectile is a significant challenge~\cite{Moore1968,Watson1970_1}. Moreover, very large pressures are required to reach the implosion velocities needed to prevent the driver gas from leaking past the explosive pinch due to the high driver gas pressures in the launch tube and the development of a boundary layer behind the projectile~\cite{Baum1973_1,Baum1973_2}. Although it is possible to achieve such implosion velocities via the impact of explosively driven flyers, the cumulative jet that results from the implosion may threaten to damage the projectile~\cite{Baum1973_1,Baum1973_2}. 

\begin{figure}[tb]
	\centering
	\includegraphics[width=1.0\columnwidth]{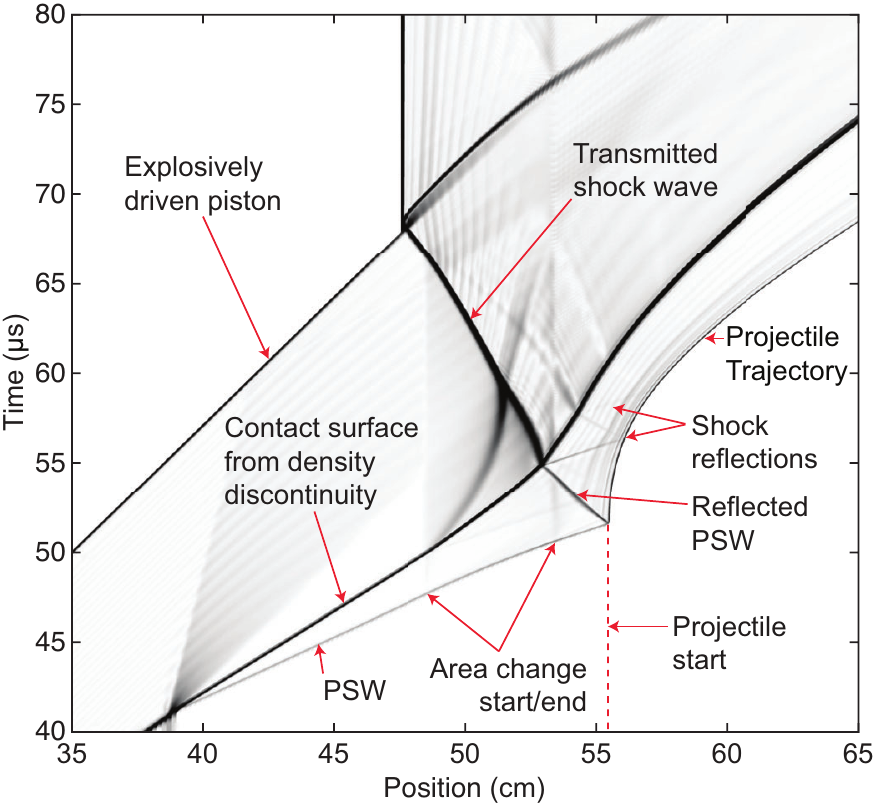}
	\caption{Labelled schlieren position--time plot showing the internal gas flow for a launcher with a pressure discontinuity in the driver. The density gradients show the shock-wave reflections responsible for maintaining a high driving pressure on the projectile.}
	\label{Fig:twopressureschlieren}
\end{figure}

A more practical approach may be to modify the design of the driver such that the launch cycle maintains a high driving pressure on the projectile. As will be shown, this can be accomplished by having a discontinuity in the initial density of the driver gas, meaning that either the molecular weight or pressure of the gas at the start of the driver is larger than the gas near the projectile. The initial propellant density at the start of the driver can be set such that the driver operates near its limiting pressure, which corresponds to a post-shock pressure of approximately \SI{1.2}{\giga\pascal}, while the gas near the projectile is at a much lower density in order to maintain a reflected shock pressure that is near that of the standard launcher. The initial contact surface (density discontinuity) travels with the flow and acts much like a piston: shock waves formed by the reflection of the PSW from the area change section and the projectile are partially reflected by the high impedance contact surface towards the projectile, which re-compresses the expanding flow. The operation of the density discontinuity launch cycle is depicted in the schlieren position--time plot of Figure~\ref{Fig:twopressureschlieren}, which was produced using an ideal IDL simulation in which there was a pressure discontinuity in the driver. The density gradients in the flow show the shock reflections responsible for maintaining a high driving pressure on the projectile.

\begin{figure}[ht]
	\centering
	\includegraphics[width=1.0\columnwidth]{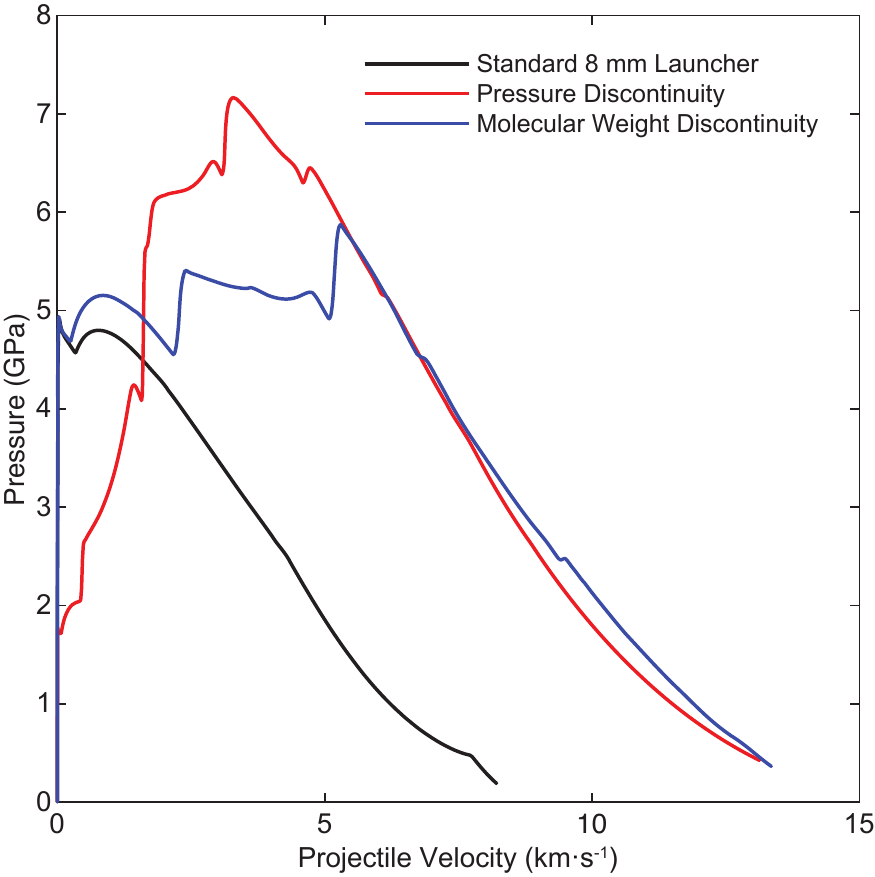}
	\caption{Projectile driving pressure as a function of velocity for internal ballistics simulations comparing the standard \SIadj{8}{\milli\metre} design presented in this work to two variations of the density discontinuity launcher technique.}
	\label{Fig:advancedconceptssim}
\end{figure}

The potential benefit of the density discontinuity technique is best illustrated by observing the evolution in the projectile driving pressure as a function of velocity using simulations. Figure~\ref{Fig:advancedconceptssim} shows the predicted evolution in driving pressure for two variations of the density discontinuity technique along with that of the standard \SIadj{8}{\milli\metre} launcher. All three simulations were performed with the ideal solver (impermeable piston, rigid launcher walls, no ablation) and the geometry of the \SIadj{8}{\milli\metre} launcher presented in this work. The “two-pressure launcher” simulation had helium at a pressure of \SI{8.8}{\mega\pascal} at the start of the driver and \SI{0.5}{\mega\pascal} on the projectile side, with the initial discontinuity located at 24 inner diameters (\SI{39}{\centi\metre}) from the start of the driver. The “molecular weight discontinuity” simulation had a helium-argon mixture of 60\% by volume helium (mixture molecular weight of \SI{18.4}{\kilo\mole\per\kilo\gram}) at the start of the driver and pure helium near the projectile, both at a pressure of \SI{2.1}{\mega\pascal} and with the discontinuity located at 22 inner diameters (\SI{36}{\centi\metre}) from the start of the driver. The initial conditions in both cases ensured that the pressure in the driver never exceeded \SI{1.1}{\giga\pascal}, which is below its established limit of operation. The traces of Figure~\ref{Fig:advancedconceptssim} clearly demonstrate that the interactions between the contact surface and shock waves in the flow maintain a high driving pressure on the projectile. This results in a much more efficient launch cycle where the average driving pressure on the projectile is significantly increased without increasing the initial shock loading. In particular, a driver with a pressure discontinuity can be tailored to give a much lower reflected shock pressure (\SI{1.8}{\giga\pascal} in this simulation), while allowing the pressure to ramp up beyond \SI{5}{\giga\pascal} in order to reach the velocities of interest. The reduction in the strength of the shock wave transmitted into the projectile is expected to significantly reduce the likelihood of projectile damage. As can be seen in Figure~\ref{Fig:advancedconceptssim}, both density discontinuity techniques may allow for a projectile velocity increase of several km/s using the standard \SIadj{8}{\milli\metre} IDL geometry simply by modifying the initial conditions in the driver. 

\begin{figure*}[t]
	\centering
	\includegraphics[width=1.0\textwidth]{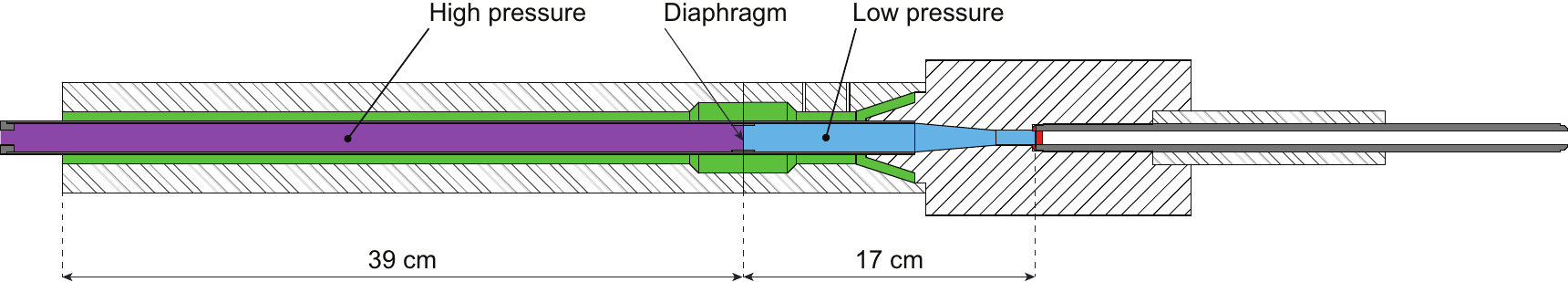}
	\caption{Schematic of the design of the two-pressure launcher used in the experiments presented in Figure~\ref{Fig:twopressurepdv}.}
	\label{Fig:TwoPressureSchematic}
\end{figure*}

\begin{figure}[htb]
	\centering
	\includegraphics[width=1.0\columnwidth]{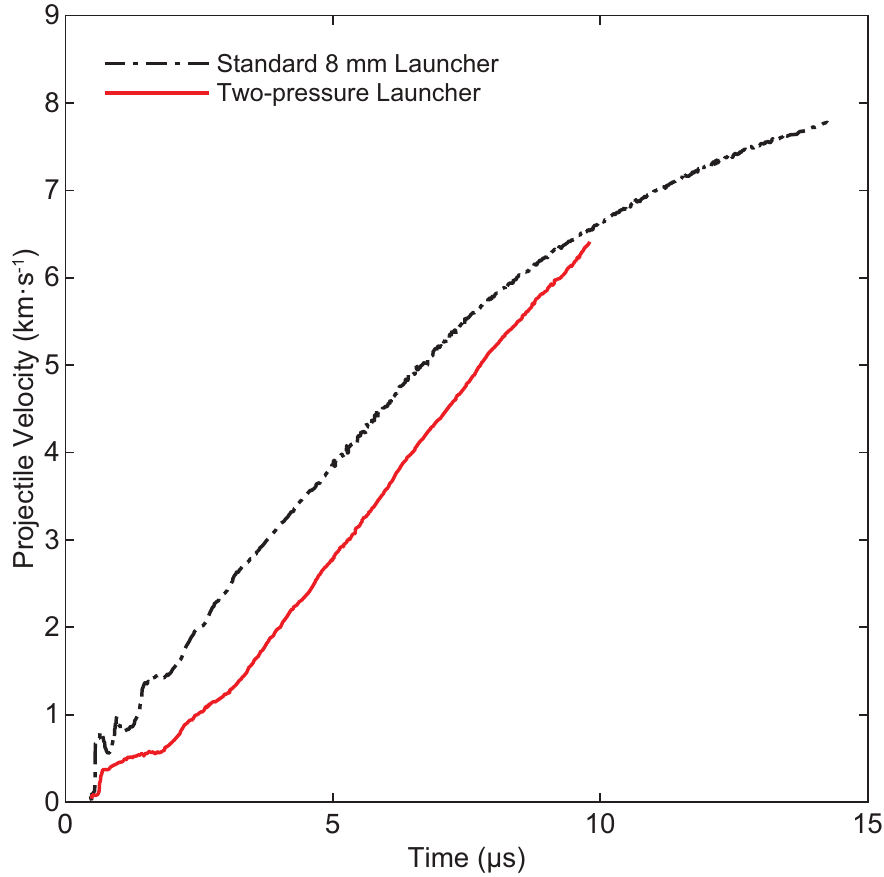}
	\caption{Results of down-bore velocimetry experiments performed on the \SIadj{8}{\milli\metre} implosion-driven launcher. The performance of the standard \SIadj{8}{\milli\metre} launcher is compared to that of a two-pressure launcher.}
	\label{Fig:twopressurepdv}
\end{figure}

Although the density discontinuity launch cycle should be relatively easy to implement, creating the initial conditions in the driver presents a challenge. In the case of a pressure discontinuity, a diaphragm must be used to hold the initial pressure differential (approximately \SI{8}{\mega\pascal}). The pressures and flow velocities behind the PSW are sufficiently large that the diaphragm will shear off upon impact of the PSW, much like a very thin projectile. The finite time needed to accelerate the diaphragm will disrupt the flow and cause the PSW to partially reflect, which will generate large pressures that could cause significant loss of driver gas, failure of the detonation front, or pre-detonation of the explosive. Moreover, creating a mechanical assembly that holds the diaphragm in place and seals the driver gas while not disrupting the implosion of the driver tube is troublesome.

The advantage of varying the molecular weight of the gas is that the density discontinuity can be established without a pressure gradient so that a very thin diaphragm could be used or be removed altogether. For example, by extending the driver tube well beyond the start of the explosives, it would be possible to arrange an auxiliary shock tube where the initial conditions are such that the desired gas distribution (pressure and contact surface location) is achieved once the wave dynamics have decayed and the gasses are stationary. The shock tube diaphragm would be located before the start of the explosives, such that it does not disturb the operation of the driver. Although there would certainly be some smearing of the contact surface during the shock tube flow, the main features of the launch cycle described above would still be present. More importantly, without the diaphragm the driver would operate exactly as it would in a standard launcher, eliminating any possibility of driver gas loss or disruption of the explosively driven piston.

Preliminary experiments were performed to attempt the pressure discontinuity technique on launchers with the same geometry and projectile design as the \SIadj{8}{\milli\metre} IDL presented in Section~\ref{Sec:DetailedDesign}. The two-pressure launcher and standard launcher were compared with down bore velocimetry experiments and high-speed photography experiments that measured the exit velocity and condition of the projectile. A schematic of the two-pressure launcher can be seen in Figure~\ref{Fig:TwoPressureSchematic}. The driver was filled with helium at \SI{9.0}{\mega\pascal} in the high-pressure section and \SI{1.0}{\mega\pascal} near the projectile. A \SIadj{0.25}{\milli\metre}-thick 7075 aluminum diaphragm separated the two sections and was coupled to the driver with threads that were sealed with epoxy. The diaphragm was located \SI{39}{\centi\metre} (24 inner diameters) from the start of the driver. The thickness of the explosive layer near the diaphragm was increased from \SI{5.6}{\milli\metre} to \SI{10.8}{\milli\metre} in order to accommodate the expected increase in driver-tube expansion and prevent failure of the detonation front. All other launcher parameters were identical to the \SIadj{8}{\milli\metre} launcher presented in Section~\ref{Sec:DetailedDesign}. The parameters for the standard launcher were the same as those presented in Table~\ref{tab:8mmResults}.

The results of down-bore velocimetry experiments comparing the early projectile acceleration history of the two-pressure launcher with that of the standard \SIadj{8}{\milli\metre} IDL is presented in the velocity--time plot of Figure~\ref{Fig:twopressurepdv}. The experimental arrangement for the PDV down bore experiments was nearly identical to that of Section~\ref{Sec:DownBore}, with the only notable difference being that the PDV probe holder was sealed in order to draw vacuum on the launch tube. As can be seen in Figure~\ref{Fig:twopressurepdv}, the strength of the shock wave initially transmitted into the projectile for the pressure gradient launch cycle was significantly less than for the standard launcher. As expected, the projectile base pressure for the pressure gradient launcher gradually increased to the point where acceleration exceeded that of the standard launcher. Unfortunately, the PDV signal was lost early in the launch cycle (at \SI{5.5}{\kilo\metre\per\second}), so it is not possible to examine how the projectile acceleration was affected later in the launch cycle. The muzzle velocity of the pressure gradient launcher was measured to be \SI{8.4}{\kilo\metre\per\second} in a follow-on experiment with high-speed videography, with the projectile being in good condition. This velocity is significantly less than the results for the standard \SIadj{8}{\milli\metre} launcher presented in Table~\ref{tab:8mmResults} (\SI{10.4}{\kilo\metre\per\second} and \SI{10.2}{\kilo\metre\per\second}) and suggests that the presence of a diaphragm affected the launch cycle. Indeed, preliminary internal ballistics simulations that included the mass of the diaphragm (treated as a lumped mass) have shown that the diaphragm has a significant effect on the internal ballistics and ultimately leads to lower projectile exit velocities~\cite{HildebrandThesis}. This effect, combined with the expected disruption of the driver due to the large transient pressures near the diaphragm are the most likely cause of the unexpectedly low projectile exit velocity.

Preliminary tests have clearly demonstrated the ability of the density gradient launch cycle to apply a more gentle initial shock loading on the projectile while allowing the pressure to progressively increase such that the projectile acceleration matches or exceeds that of the standard launcher. However, these tests have revealed the need for minimizing the mass of the diaphragm or eliminating it entirely in order to achieve the full potential of the density discontinuity launch cycle. Manipulating the molecular weight of the gas rather than the initial pressure in order to establish a density discontinuity in the driver would significantly reduce the thickness requirement of the diaphragm or even eliminate the need for one altogether. This approach is seen as the most promising strategy for achieving the potential improvements in projectile velocity that the simulations using the internal ballistics solver suggest are possible.

\end{appendices}

\end{document}